\documentclass[twocolumn, twocolappendix, usenatbib, useAMS]{aastex631}
\usepackage{times, amsmath, amssymb, amsfonts, latexsym}
\usepackage{fleqn, multirow, graphicx, color, enumitem}
\usepackage{array, mathtools,subfigure, esint, CJK}
\usepackage{soul, hyperref}
\newcommand{\uat}[2]{\href{http://astrothesaurus.org/uat/#2}{#1 (#2)}}

\def\mpch{\,{h^{-1} {\rm Mpc}}}          \def\hmpc{\,{h {\rm Mpc}^{-1}}}
\def\dd{{\rm d}} 
\makeatletter
\@for\next:={int,iint,iiint,iiiint,dotsint,oint,oiint,sqint,sqiint,
	ointctrclockwise,ointclockwise,varointclockwise,varointctrclockwise,
	fint,varoiint,landupint,landdownint}\do{%
	\expandafter\edef\csname\next\endcsname{%
		\noexpand\DOTSI
		\expandafter\noexpand\csname\next op\endcsname
		\noexpand\ilimits@
	}%
}
\makeatother
\shorttitle{Identifying 2D Halos Using Continuous Wavelet Transform}
\shortauthors{Li, Wang \& He}
\begin{document}
\begin{CJK*}{UTF8}{gbsn}
\title{Identifying Halos in Cosmological Simulations with Continuous Wavelet Analysis: The 2D Case}

\author[0009-0003-1625-8647]{Minxing Li (李敏行)}
\affiliation{College of Physics, Jilin University, Changchun 130012, P.R. China.}

\author[0000-0003-4064-417X]{Yun Wang (王云)}
\affiliation{College of Physics, Jilin University, Changchun 130012, P.R. China.}

\author[0000-0001-7767-6154]{Ping He (何平)}
\affiliation{College of Physics, Jilin University, Changchun 130012, P.R. China.}
\affiliation{Center for High Energy Physics, Peking University, Beijing 100871, P.R. China.}

\correspondingauthor{Ping He}
\email{hep@jlu.edu.cn}
\begin{abstract}
Continuous wavelet analysis is gaining popularity in science and engineering for its ability to analyze data across both spatial and scale domains simultaneously. In this study, we introduce a wavelet-based method for halo identification and assess its feasibility in two-dimensional (2D) scenarios. We begin by generating four pseudo-2D datasets from the SIMBA dark matter simulation by compressing thin slices of three-dimensional (3D) data into 2D. Subsequently, we calculate the continuous wavelet transform (CWT) directly from the particle distributions, identify local maxima that represent actual halos, and segment the CWT to delineate halo boundaries. A comparison with the traditional friends-of-friends (FOF) method shows that while our CWT-identified halos contain slightly fewer particles, they have smoother boundaries and are more compact in dense regions. In contrast, the CWT method can link particles over greater distances to form halos in sparse regions due to its spatial segmentation scheme. The spatial distribution and halo power spectrum of both CWT and FOF halos demonstrate substantial consistency, validating the 2D applicability of CWT for halo detection. While our identification scheme has been tested solely in a limited 2D context and has shown some performance limitations, its linear time complexity of $\mathcal{O}(N)$ and consistency with the FOF method suggest its suitability for analyzing significantly larger datasets in the future and its potential to be extended to the 3D case.
\end{abstract}
\section{Introduction}
\label{sec:intro}

In modern cosmology, it is assumed that our Universe was highly homogeneous in its early stages. Over time, these initial minute density perturbations evolved into a variety of structures, such as dark matter halos, galaxies, clusters, and voids. This evolution occured through both linear and nonlinear processes, driven by long-range gravity \citep{Mo2010}. To study the Universe's evolution, it is essential to conduct observations of astronomical objects at high redshifts and over vast cosmic distances. These observations are complemented by numerical simulations, recognized as indispensable tools for cosmological research. Such simulations begin with initial conditions and then solve the $N$-body dynamic equations or hydrodynamic equations, thereby modeling the Universe's evolution. Within these simulations, the density of the Universe is usually represented by particles, typically dark matter and baryonic matter particles (in the case of Lagrangian codes) \citep{Vogelsberger2020}, which perform a Poisson sampling of the simulated density field \citep{Peebles1980}.

With the rapid advancement of computer science, cosmological simulations have progressed from rudimentary $N$-body models involving pure dark matter bound by gravity \citep{Press1974} to sophisticated adaptive moving mesh magnetohydrodynamic simulations \citep{Nelson2019} and meshless finite mass hydrodynamic simulations \citep{Dave2019}, which incorporate the processes of feedback and interactions. These advanced simulations enable us to model the evolution of the Universe from redshifts of a few hundred to the present day, within a simulation box spanning millions or even billions of parsecs in comoving units. While the simulations align with observations in various aspects - such as the existence of the cosmic web, the characteristics of stars, and the gas content of galaxies - the analysis of simulated particle data remains challenging. For instance, when determining statistical properties like the power spectrum, particles are commonly assigned to a regular grid using window functions, such as the cloud-in-cell (CIC) or triangular-Shaped cloud methods \citep[see][]{Hock-East1981}. Moreover, constructing a halo catalog to study the clustering of matter in the simulated universe necessitates identifying individual halos at the particle data level and considering a cluster of particles as a single halo.

For this purpose, we can employ various halo finders, such as the Amiga Halo Finder \citep{Knollmann2009}, the adaptive spherical overdensity halo finder \citep{Planelles2010, Valles-perez2022}, HOP \citep{Eisenstein1998}, and ROCKSTAR \citep{Behroozi2013}. Halo finders can generally be categorized into two types: density peak locators and particle collectors. The former, based on the spherical-overdensity method \citep{Press1974}, identifies the density peaks and subsequently selects particles encircling these peaks to constitute halos. The latter, exemplified by the friends-of-friends (FOF) method \citep{Davis1985}, links particles that satisfy specific conditions together (e.g., the distance below the linking length in FOF, or toward the densest direction in HOP).

Generally speaking, a halo finder typically involves the following processes: identifying halo candidates and excluding spurious halos, collecting particles associated with these candidates, determining the halo center, conducting a self-boundness check and unbinding free particles, delineating the halo boundary, and tracing halos across different snapshots. Due to the extreme variability among different halo identification methods, not all halo finders undertake each of the aforementioned processes. For instance, some FOF-based halo finders do not generate a halo candidate catalog. The standard FOF halo finder excludes the unbinding process, and most early halo finders do not track the time evolution of halos. Despite the existence of these disparate halo finders, a degree of uncertainty persists. A primary source of uncertainty lies in defining the halo boundary, which is closely related to the fundamental question of what constitutes a halo. The criteria for this definition vary significantly across different halo finders \citep{Knebe2011, Knebe2013}.

Among them, the most widely used algorithm is the FOF method, which calculates a linking length based on the particle number density and then connects particles that are closer than this length. Despite its simplicity, this method is popular for identifying dark halos from particle data, as it offers a reliable measure of halo distribution. However, the FOF method has limitations in that it has a time complexity of $\mathcal{O}(N\log N)$ when utilizing a minimal spanning tree for organizing particle data, and it employs an artificial parameter $b$ in the linking length. Worst of all, it lacks a clear boundary definition, leading to the `linking bridges' problem, erroneously connecting two distinct halos if there is a chain of closely spaced particles (below the linking length) between them \citep{Springel2001}. 

In addition to the FOF method, the wavelet transform method, first introduced by \citet{Grossmann1985}, offers an alternative approach to clustering analysis. This technique is extensively utilized in astronomy and cosmology to identify structures, such as analyzing clusters in X-ray data \citep[e.g.,][]{Vavilova2018, Freeman2002}, infrared data \citep{Wang2008}, and gamma-ray data \citep{Ciprini2007}. Additionally, it serves to investigate how discreteness effects in $N$-body simulations affect the statistical diagnostics of the cosmological density field \citep{Romeo2008}, trace the time evolution of the cosmic web \citep{Einasto2011}, identify sunspots \citep{Djafer2012}, and characterize coronal mass ejections \citep{Gonzalez2010}. Furthermore, the wavelet method is employed to detect baryon acoustic oscillation structures \citep{Arnalte-Mur2012}, locate ridges of spiral arms \citep{Patrikeev2006}, and identify point source contamination in cosmic microwave background observations \citep{Cayon2000}.

There are two distinct wavelet methods: the discrete wavelet transform (DWT) and the continuous wavelet transform (CWT). The DWT decomposes a signal into a set of orthogonal wavelet bases, offering a compact representation. In contrast, the CWT involves the convolution of the signal with a wavelet that is scaled by a mother wavelet. The CWT computation is typically performed on a grid, or on image pixels in observations. It employs the fast Fourier transform and exploits the property that convolution in real space is equivalent to multiplication in Fourier space \citep[e.g.,][]{Arnalte-Mur2012, Grebenev1995, Barnard2004, Kazakevich2004}.

In structure identification, the DWT is frequently implemented via the `$\Grave{{\rm a}}$ trous' algorithm. This algorithm filters an image with scaling functions at doubled magnification scales and then decomposes it into wavelets at various scales, along with a residual \citep[e.g.,][]{Bijaoui1991, Pagliaro1999, Ellien2021, Lazzati1999, Moretti2004}. This algorithm, distinct from the standard DWT, employs a nonorthogonal wavelet basis, enhancing its ability  for structural localization \citep{Chereul1998}. It enables rapid computation of grid DWTs at multiple scales with a constant resolution, but the scaling is fixed to a doubly magnified set of scales. 

Compared with the DWT, the CWT typically demands more time and CPU resources. However, when dealing with discrete particle data from simulations, an alternative method exists to compute the CWT with a time complexity of $\mathcal{O}(N)$. This approach views the density field as a sum of Dirac delta functions, meaning that the corresponding CWT is the convolution of these Dirac functions with a wavelet, yielding a sum of wavelets. Although some authors describe this method as a DWT \citep[e.g.,][]{Bendjoya1991, Slezak1993, Flin2006}, the outcome of this convolution is inherently continuous, as its value can be determined at any position and scale, regardless of resolution.

In this paper, we introduce a novel wavelet-based method for two-dimensional (2D) halo identification, visualize the results, and compare them with the traditional FOF method. The paper is organized as follows. In Section~\ref{sec:frame}, we provide a brief overview of the CWT method, the FOF method, the utilized SIMBA dataset, and the derived pseudo-2D dataset. Section~\ref{sec:halo-ident} details our multiscale CWT halo identification approach. In Section~\ref{sec:resolution}, we calculate the statistical significance and investigate the impact of resolution on our method. Section~\ref{sec:result} presents our identification findings and compares them with the FOF halo identifications. Finally, we offer conclusions and discuss the implications of our results in Section~\ref{sec:concl}. The actual performance of our current program is presented in the Appendix~\ref{sec:performance}.

\section{Theoretical Foundations and Methods}
\label{sec:frame}
\subsection{The CWT}

All the wavelets must meet two criteria: the admissibility condition and the square integrability condition. The former stipulates that the wavelet's energy must attenuate to zero as the independent variable approaches infinity, while the latter mandates that the integral of the wavelet's square over the entire space be finite. The CWT is defined as the convolution of the signal $f(\mathbf{x})$ with a wavelet $\Psi (w,\mathbf{x})$ (assuming both the signal and the wavelet are real) \citep{Slezak1990},
\begin{equation}
    \tilde{f}(w,\mathbf{x}) =\int_{-\infty}^{\infty}f (\mathbf{u})\Psi (w,\mathbf{x}-\mathbf{u}) \dd\mathbf{u},
\end{equation}
where $\Psi (w,\mathbf{x})$ is a scaled version of the so-called mother wavelet $\Psi (\mathbf{x})$, as follows:
\begin{equation}
    \Psi (w,\mathbf{x}) = w^{d/2}\Psi (w\mathbf{x}),
    \label{scaled_wavelet}
\end{equation}
with $d$ denoting the dimension of space \citep{Wang2022a}. In our 2D case, Equation~(\ref{scaled_wavelet}) simplifies to
\begin{equation}
    \Psi (w,\mathbf{x}) = w\Psi (w\mathbf{x}).
\end{equation}
We introduce the scale parameter $w$ instead of the traditional scale indicator $a$, with the relationship between them being $w=1/a$ \citep{Wang2022a,Wang2022b}. For discrete particle data from simulations, their density field can be expressed as the sum of the Dirac delta functions $\delta^{(\rm D)}(\mathbf{x})$ located at their positions and weighted by their masses:
\begin{equation}
    \rho(\mathbf{x})=\sum_{i}m_i\delta^{(\rm D)}(\mathbf{x}-\mathbf{x}_i),
\end{equation}
where $m_i$, $\mathbf{x}_i$ represent the mass and position vector of the $i$ th particle, respectively. Thus, the CWT of the particle density field is the convolution of the Dirac function with our wavelet, which is simply the translated wavelet itself
\begin{flalign}
\label{CWT_Analysis}
  \tilde{\rho} (w,\mathbf{x}) = &\int_{-\infty}^{\infty}\rho (\mathbf{u})\Psi (w,\mathbf{x}-\mathbf{u})\dd\mathbf{u} \nonumber \\
  = & \int_{-\infty}^{\infty}\sum_{i}m_i\delta^{(\rm D)}(\mathbf{u}-\mathbf{x}_i)\Psi (w,\mathbf{x}-\mathbf{u})\dd\mathbf{u}  \nonumber \\
  = & \sum_{i}m_i\int_{-\infty}^{\infty}\delta^{(\rm D)}(\mathbf{u}-\mathbf{x}_i)\Psi (w,\mathbf{x}-\mathbf{u})\dd\mathbf{u} \nonumber \\
  = & \sum_{i}m_i\Psi (w,\mathbf{x}-\mathbf{x}_i).
\end{flalign}
It is almost impossible to analyze Equation~(\ref{CWT_Analysis}) analytically. In this context, we must analyze millions of functions simultaneously (the number of functions equals the number of particles). Instead, we discretize Equation~(\ref{CWT_Analysis}) onto a regular grid; that is, we calculate the value of Equation~(\ref{CWT_Analysis}) at specific grid points $\mathbf{x}_{\rm g}$ to represent it: 
\begin{equation}
    \tilde{\rho} (w,\mathbf{x}_{\rm g}) = \sum_{i}m_i\Psi (w,\mathbf{x}_{\rm g}-\mathbf{x}_i),
    \label{CWT} 
\end{equation} 
which is much more manageable.

In this work, we utilize 2D isotropic Mexican hat (MH) wavelet, defined as
\begin{equation}
    \Psi (w,\mathbf{x}) = w(2 - w^2 r^2)e^{-\frac{w^2 r^2}{2}},
\end{equation}
which is derived from its mother wavelet
\begin{equation}
\label{eq:mhw}
\Psi (r) = (2-r^2)e^{-\frac{r^2}{2}},
\end{equation}
where $r=|\mathbf{x}|$ denotes the magnitude of the vector $\mathbf{x}$. This wavelet can be considered, apart from a prefactor, as the first derivative of the Gaussian function with respect to $w$ \citep[see][for details]{Wang2021}.

Even when employing isotropic wavelets, they can still be used to detect anisotropic structures \citep{Vavilova2018}. To simplify our calculations, we opt for isotropic wavelets with compact support, or at least with an approximate compact support. Furthermore, if a wavelet is not orthogonal, a particular structure might exhibit local maxima at several scales, rather than solely at the scale that most closely matches the structure's size. This could lead to ambiguity when identifying halos within a multiscale analysis \citep{Hayn2012, Ellien2021}. However, it is not feasible for a wavelet to simultaneously possess compact support, isotropy, and orthogonality \citep{Daubechies1992}. Therefore, we select the MH wavelet for its isotropy and approximate compact support \citep{Wang2021}. The CWT has a notable property: the CWT of any constant or linear component of a signal is zero. This characteristic provides us with a method to distinguish between different structures, even in highly dense environments, and offers a natural delineation.

\begin{figure*}[t]
    \centering
    \includegraphics[width=0.95\textwidth]{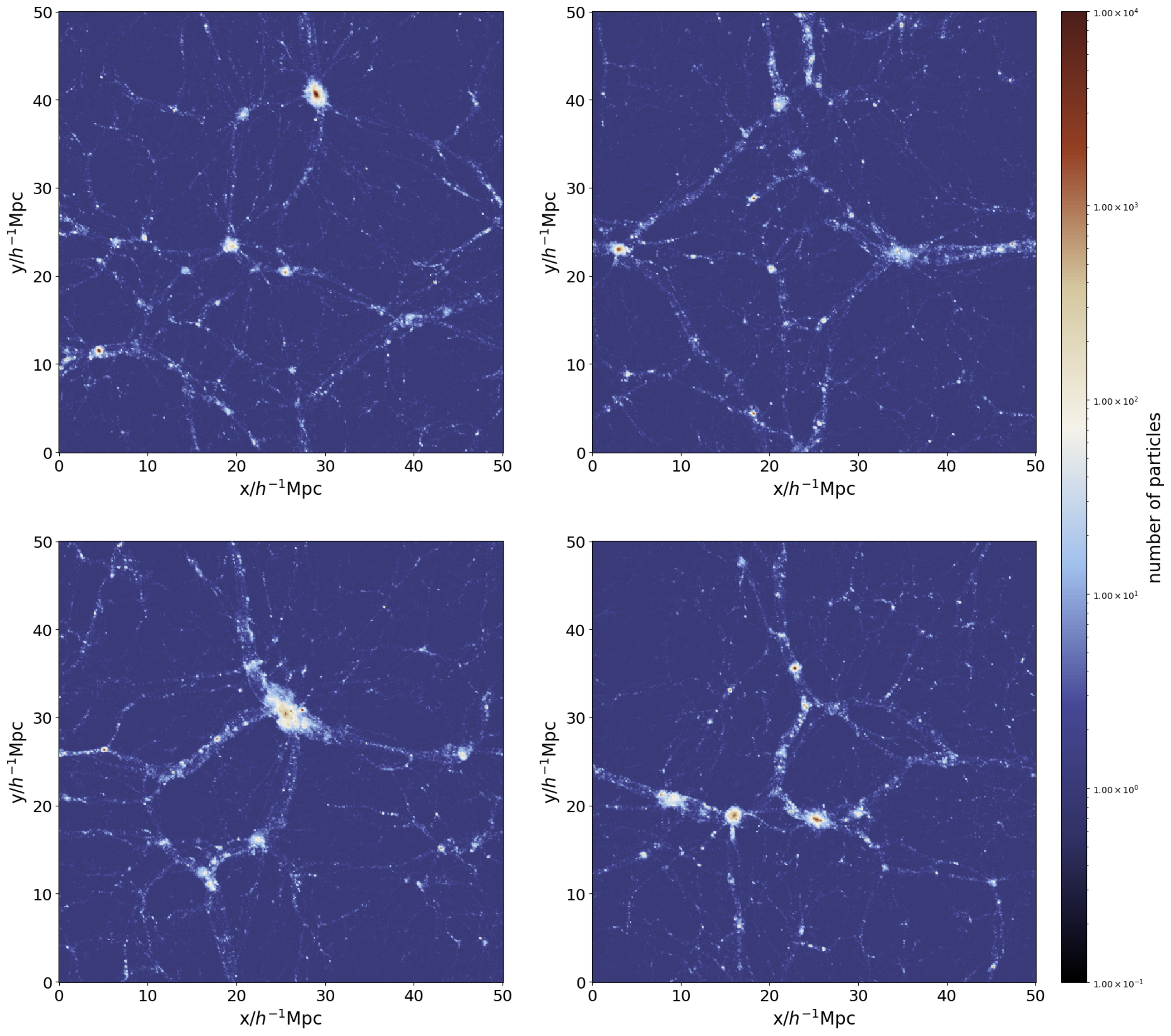}
    \caption{The density field of our pseudo-2D data in the CIC grid. From top left to bottom right are slices at ${\rm z} = 11.4$, $28.4$, $38.0$ and $45.7\mpch$, respectively. The thickness of all the four slices is $\Delta{\rm z} = 0.1\mpch$ and all the grids are made of $512^2$ cells.}
    \label{CIC_gird}
\end{figure*}

\begin{deluxetable*}{lcc}
\label{tab:parameter}
\tablehead{
\colhead{Parameter} & \colhead{Meaning} & \colhead{Definition} 
}
\tablecaption{List of the parameters used in This paper, along with Their Meanings and Definitions.}
\tablenum{1}
\startdata
        a      &   The physical scale of the wavelet  &  None                   \\
        w      &   The scale parameter                &  $1/a$                  \\
        $k_w$  &   The dimensionless scale parameter     &  $\frac{L_{\rm box}}{1000} w$         \\
        $N_{\rm g}$  &   The resolution of the grid      & ${\rm Int}(1000\times k_w) + 500$     \\ 
	  $N_{\rm W}$  &   The resolution of the wavelet   & ${\rm Int}(2 \times(10/k_w + 10)) + 1$\\
        $N_{\rm cs}$ &   The resolution of cross-scale comparison  &   1000                      \\
        $N_{\rm th}$ &   The least particle number in halo         &   5                         \\
\enddata
\end{deluxetable*}

\subsection{Friends-of-Friends Algorithm}

The FOF method categorizes particles into distinct halos \citep{Davis1985}. The grouping process can be summarized as follows: First, an arbitrary particle is selected. Then, this algorithm searches for its nearby particles and links any particle whose distance to the selected one is less than the linking length, $l_{\rm link}$; these are termed the friends of the initial particle. Subsequently, the search continues for the friends of each newly linked particle. This linking continues to include additional friends whose distance to any current friend is also less than $l_{\rm link}$. The process iterates to identify friends of friends until no further particles meet the linking criterion. These interconnected particles ultimately form a FOF group.

After that, select another arbitrary particle that is not already in any group and search for its friends to form a new group. Repeat this process until all particles are assigned to groups, even if some groups contain only one particle. Finally, exclude groups that have fewer than a specified threshold number of particles, $N_{\rm th}$; for example, $N_{\rm th}=5$. This results in the classification of halos comprising five or more particles.

Obviously, there are two artificial parameters in FOF, $l_{\rm link}$ and $N_{\rm th}$. The former is usually defined as
\begin{equation}
    l_{\rm link} = b\left(\frac{V}{N}\right)^{1/3},
    \label{linking_length}
\end{equation}
where $V$ denotes the volume of the simulation box, $N$ is the number of particles, and $b=0.2$ so that the average density in the halo is 180 times the average density in the box \citep{More2011}. In the 2D case, Equation~(\ref{linking_length}) can be rewritten as
\begin{equation}
    l_{\rm link} = b\left(\frac{S}{N}\right)^{1/2},
\end{equation}
where $S$ is the area of the simulation box and we still set $b=0.2$. However, the choice for the latter, $N_{\rm th}$, is much more arbitrary. Although $N_{\rm th}$ represents the mass of the smallest halo, there is no consensual definition\footnote{Some authors set $N_{\rm th}=32$ (see \url{http://swift.dur.ac.uk/docs/FriendsOfFriends/algorithm\_description.html}), while others have different definitions (e.g., ytastro toolkit has the default $N_{\rm th}=5$) \citep{Turk2011}.}.

\subsection{SIMBA Simulation and Pseudo-2D Data}

SIMBA \citep{Dave2019} is a meshless finite mass hydrodynamic cosmological simulation based on the GIZMO cosmological gravity plus hydrodynamics solver \citep{Hopkins2015}, assuming $h=0.68$, $\Omega_{\rm \Lambda } = 0.7$, $\Omega_{\rm m} = \Omega_{\rm dm} + \Omega_{\rm b}=0.3$, $\Omega_{\rm b}=0.048$, $n_s = 0.97$, and $\sigma_8 = 0.82$, which are the Planck cosmological parameters \citep{Planck2016}. This simulation consists of several sub-simulations with different box lengths, particle numbers, and even different physics, e.g., no active galactic nuclei, no jet, and pure dark matter.

This paper aims to identify halos within the particle data, without delving into the detailed physics of the simulation. We therefore use the {\tt m50n512} dark matter sub-simulation sample, which is a pure dark matter simulation devoid of complicated processes such as star formation, gas cooling, and feedback\footnote{This dataset was downloaded from \url{http://simba.roe.ac.uk/simdata/m50n512/dm/snapshots/snap_m50n512_151.hdf5}, but access to this sub-simulation is currently disabled.}. This sub-simulation has a box length of $50h^{-1}\rm Mpc$ and contains $512^3$ dark matter particles, with a mass resolution of $9.6\times 10^7 M_\odot$.

To obtain 2D data, we select the snapshot at redshift $z=0$ from this sub-simulation and cut a slice in the z-direction, keeping only the x, y coordinates of these particles. Though the dimension is reduced from the original three-dimensional (3D) data, this pseudo-2D dataset retains substantial structural information, proving adequate for testing our halo identification algorithm.

We use four slices of pseudo-2D data at ${\rm z} = 11.4$, $28.4$, $38.0$ and $45.7\mpch$ with a thickness of $\Delta{\rm z} = 0.1\mpch$, containing 323,716, 228,187, 303,692, and 312,372 particles, respectively. The CIC grids of these slices, each consisting of $512^2$ cells, are shown in Figure~\ref{CIC_gird}.

\section{Halo Identification}
\label{sec:halo-ident}

\subsection{Procedure of Identification}
We provide a detailed description of our halo identification algorithm, which consists of the following steps:

(1) Calculation of the grid CWT at different scales using a grid-assigning method, as shown in Equation~(\ref{CWT}), which results in an $\mathcal{O}(N)$ time complexity. 

(2) Determination of thresholds to filter those local maxima generated by Poisson noise at different scales, as discussed in Section~\ref{sec:cs-locmx}, is performed using either the Monte Carlo method or a pre calculated empirical formula.

(3) Identification of local maxima in the CWT that exceed the threshold at each scale, followed by a comparison across scales to ensure that only the largest CWT value remains at any given location.

(4) Preservation of these local maxima is accompanied by the segmentation of the grid CWT at each scale, with boundaries defined by grid points that are less than or equal to zero and exhibit a decreasing CWT value, which naturally delineates the halo's boundary.

(5) Evaluation of each maximum to confirm its location within a structure at its corresponding scale, discarding any structure that lacks a maximum.

(6) Utilizing the zoom function from scipy.ndimage, adjust the resolution of these structures and their backgrounds to a common level and overlay them in descending order of scale size. When two structures overlap, retain only the structure originating from the larger scale. The structures that remain represent potential halos.

(7) Verification of each particle's position within a structure is conducted; if confirmed, the particle is labeled with the index of the potential halo. After all particles have been verified, any halo containing too few particles is discarded.

In the following, we present the detailed calculations for each step of the process. A summary of the parameters utilized in this study is presented in Table~\ref{tab:parameter}.

\begin{figure*}[t]
    \centering
    \includegraphics[width=0.95\textwidth]{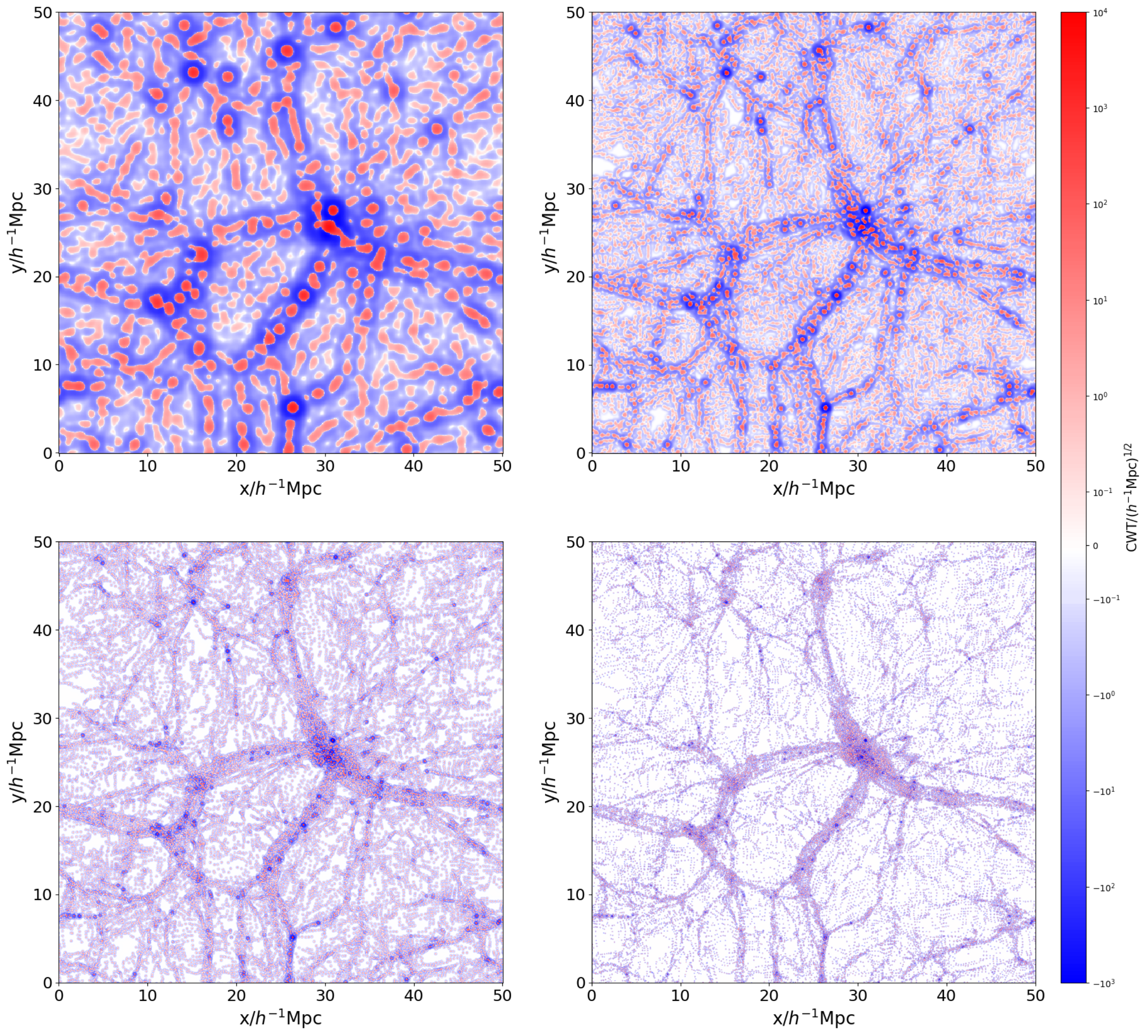}
    \caption{The 2D CWT of the slice of ${\rm z}=38\mpch$ with four different scale parameters, from top left to bottom right represent the CWT at $w = 2.29$, $5.94$, $15.75$, and $35.24\hmpc$, respectively. The sizes of the structures presented in these CWTs vary depending on the scale parameter, extending from large to small scales.}    
    \label{2dCWT}
\end{figure*} 
\begin{figure*}[t]
    \centering
    \includegraphics[width=0.95\textwidth]{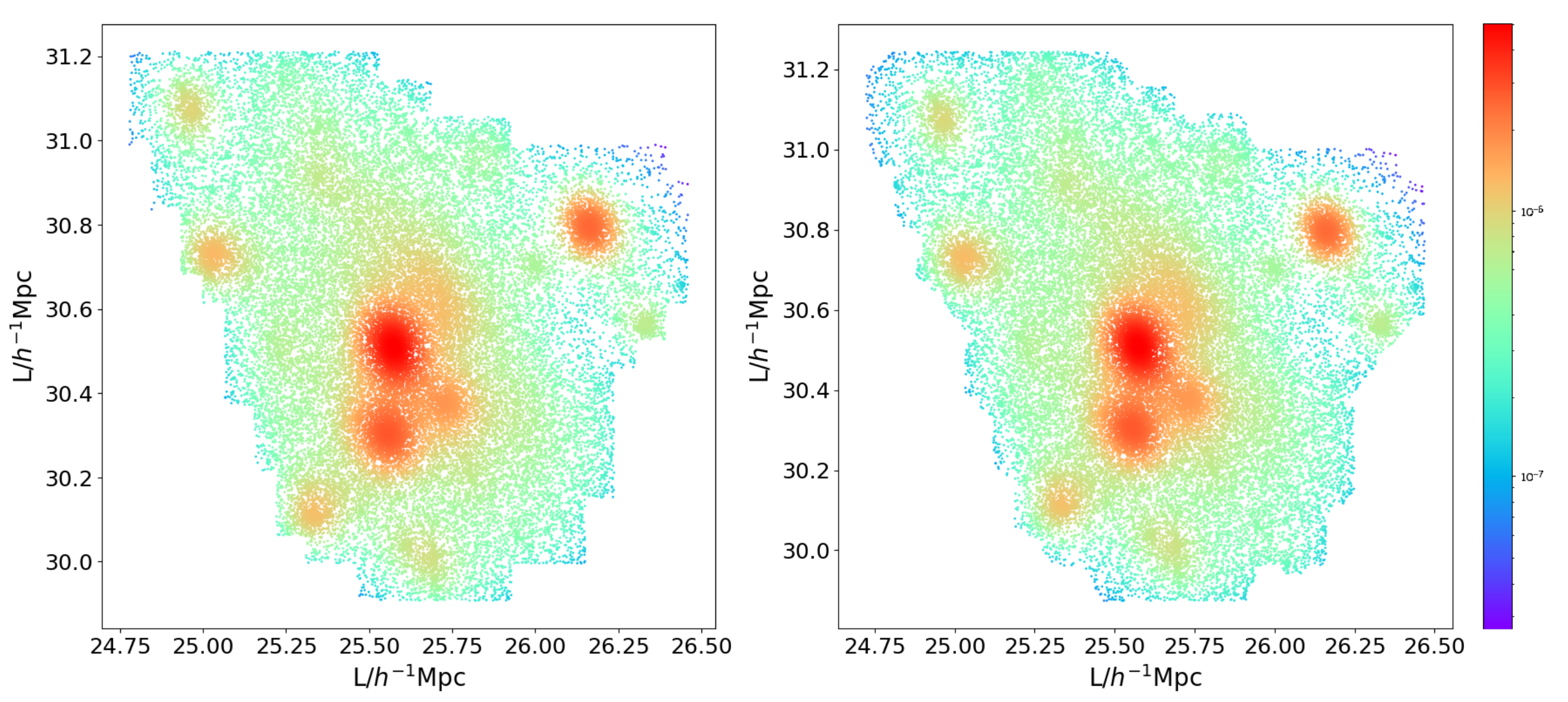}
    \caption{The same halo is depicted with directly assigned particles (left) and weighted assigned particles (right). The color represents the probability density of particles. The halo with directly assigned particles exhibits a very serrated boundary, whereas the weighted assignment results in a smoother, more rounded boundary.}
    \label{Smooth}
\end{figure*}

\subsection{Grid-based CWT Computation}
Equation~(\ref{CWT}) illustrates that the grid CWT is the sum of the wavelets centered on the position of the particles. Since the MH wavelet $\Psi(x)$ in Equation~(\ref{eq:mhw}) has an approximate compact support, it affects only a finite range near its center. This means that for some $r_{\rm c}>0$, if $|\Psi(\mathbf{x})|$ is less than, e.g., $10^{-10}$ for all the $|\mathbf{x}|>r_{\rm c}$, then we can truncate $\Psi(|\mathbf{x}|)$ to be zero wherever $|\mathbf{x}|>r_{\rm c}$. This property allows a convenient calculation of the grid CWT, since a wavelet at any scale will only have a nonzero value at grid points near its center.

We divide the simulation box into a $N_{\rm g}^2$ grid and consider only $N_{\rm W}^2$ grid points for each wavelet. The parameters are defined as follows:
\begin{flalign}
  & N_{\rm g} = {\rm Int}(1000\times k_w) + 500, \nonumber \\
  & N_{\rm W} = {\rm Int}(2 \times(10/k_w + 10)) + 1, \nonumber
\end{flalign}
where we introduce the dimensionless scale parameter $k_w$, which is expressed in terms of the length of the simulation box $L_{\rm box}$ as
\begin{flalign}
  k_w = \frac{L_{\rm box}}{1000} w.
\end{flalign}
In this way, $k_w = 1$ always represents an intermediate scale when identifying halos, and in this study, it corresponds to a scale parameter of $20\hmpc$. A dynamic mesh is created based on the variable $k_w$ to save memory and speed up computation. To accurately compute these translated wavelets on the grid, the grid point closest to the position of the particle is selected as the center of the grid wavelet, and this grid point is extended by ${\rm Int}(10/k_w) + 10$ grid points in each direction. We will discuss the resolution parameters in the above two equations in detail in the next section.

The $N_{\rm g}^2$ grid is initialized with zeros, and each individual CWT, corresponding to its particle in the $N_{\rm W}^2$ grid, is calculated and added to the appropriate part of the $N_{\rm g}^2$ grid. This part is centered at the grid point nearest to the particle and extends ${\rm Int}(10/k_w) + 10$ grid points in each direction. If a particle is located near the edge of the simulation box, its extended ${\rm Int}(10/k_w)+10$ grid points may overflow beyond the $N_{\rm g}^2$ grid. In this scenario, the overflowed grid points are relocated to the opposite side, using the periodic boundary condition applied to the particle data.


\subsection{Cross-scale Analysis of Local Maxima and Thresholding}
\label{sec:cs-locmx}

Once the CWT grid is computed, the clustering of particles can be easily analyzed \citep{Slezak1990}. When using wavelets to identify haloes, it is generally agreed that each halo should be located at the position of its corresponding local maximum. The scale at which we find the maximum represents the size of the structure \citep[e.g.,][]{Bendjoya1991, Flin2006, Grebenev1995, Slezak1993, Cayon2000, Lazzati1999, Patrikeev2006, Hayn2012}. Since our wavelet is not orthogonal, finding its local maximum directly in 3D space (2D location plus 1D-scale parameter) can be misleading, as a single structure may reach its maximum at multiple scales \citep{Hayn2012, Ellien2021}. Consequently, it is imperative to verify these maxima across scales to ensure that we identify the maximum that corresponds to the actual structure, that is, the most prominent maximum near the position of this structure.

Also, the particle distribution represents a Poisson sampling of the so-called real density field. This sampling process introduces Poisson noise that can lead to spurious local maxima when calculating the CWT using these particle data \citep{Bijaoui1992}. To mitigate this issue, a significance test should be conducted to assess the likelihood of detecting true structures as opposed to Poisson noise. We apply thresholds obtained from Monte Carlo simulations and eliminate any maxima that do not meet these thresholds across each scale. A detailed description of this process will be presented in the subsequent section.

Once we have identified all thresholded local maxima, we divide the simulation box into an additional $N_{\rm cs}^2$ grid. $N_{\rm cs}$ represents the grid resolution for the cross-scale comparison and we set its value to 1000 as summarized in Table~\ref{tab:parameter}. Due to the use of the dynamic mesh in the calculation of the grid CWT, a consistent spatial resolution is required to locate these maxima when comparing them across scales.

We then convert the location of these maxima, which is expressed as the count of grid lengths at their corresponding scale, onto the $N_{\rm cs}^2$ grid, starting from the larger to the smaller physical scales (from a smaller $w$ to a larger $w$). When we map a maximum onto the $N_{\rm cs}^2$ grid for cross-scale comparison, we compare its CWT value with the values of the surrounding $3 \times 3$ grid points centered on the grid point of this maximum. If this maximum has a higher CWT value than all nine values within the surrounding grid points, we clear the information stored in those grid points and record the information of this maximum in its corresponding grid point. This ensures that there is only one maximum with the highest CWT value within every $3 \times 3$ grid block of the $N_{\rm cs}^2$ grid. These maxima correspond to halo candidates in our method.

\subsection{Grid Segmentation and Halo Boundary Delineation}

To identify a halo, a clear boundary must be established between the halo and the background Universe. Fortunately, the wavelet method provides a natural boundary through its inherent properties. A positive CWT value is obtained in regions where the signal's shape aligns with that of the wavelet, such as at the signal's peak. Conversely, a negative CWT value is observed in regions where the signal's shape is inversely related to the wavelet, such as within the signal's valley. Additionally, the matching between the peak and the negative ring of our wavelet also leads to negative CWT values. In Figure~\ref{2dCWT} we present four different CWTs for the slice at ${\rm z} = 38.0\mpch$, covering physical scales from large to small. The result confirms the ability of the CWT to distinguish structures across scales. The CWT at a larger scale contains larger clumps, and as the scale parameter increases (toward a smaller physical scale), the size of these clumps decreases. The CWT at any given scale reflects the clustering of particles into clumps at a size corresponding to this scale, and the value of the CWT represents the amplitude or strength of its corresponding clump. 

Notably, all four CWTs exhibit a spatial distribution similar to that of the CIC density shown in Figure~\ref{CIC_gird}. Higher densities in the density field also result in higher CWT values, while the absence of particles corresponds to a CWT value approaching zero. Additionally, every positive clump in the CWT is encircled by a high-negative ring caused by the matching of the negative ring of the wavelet and the density peak. If a wavelet matches with a peak in a positive signal, it indicates that the positive part has a size similar to that of the peak, with the negative ring located outside. The location where a positive CWT value is attained is situated within the density peak, whereas the exterior portion, where the CWT value is negative, is situated outside of the density peak. In this way, points where the CWT equals zero are intermediate positions and serve as the boundary of the peak in the CWT field. However, since the CWT is calculated on a finite grid, finding the grid points where the CWT equals zero can be insufficient. In some cases, there may not be any grid points that are strictly equal to zero.

Fortunately, we can also use all the outer regions as boundaries to isolate halos. This allows us to segment the full CWT grid at each scale into `islands' of positive values surrounded by a `sea' of zero or negative values. These islands represent clusters at that scale. Similar segmentation methods have been employed in many previous studies to identify various structures within the CWT grid \citep[e.g.,][]{Baluev2020, Freeman2002}, as well as in the DWT grid \citep[e.g.,][]{Pagliaro1999, Mertens2015}.

This method proves effective for small and medium physical scales but encounters limitations at larger physical scales. The support of our wavelet is more extensive at larger scales, which results in the linking of nearby structures by `positive bridges'. Additionally, the lower grid resolution at larger scales exacerbates this situation. Regardless of the constant boundary condition applied, some structures are invariably linked by a positive bridge.

Despite the presence of substructures, a gravity-bound halo exhibits a unimodal structure. At a larger scale, these substructures have a negligible impact on the CWT. Therefore, any actual halo must have a unimodal CWT. With this in mind, we introduce a new boundary condition akin to that described in \citet{Freeman2002}, which divides closely situated structures by identifying the saddle point between them. At any scale, we start by identifying all local maxima in the CWT and use these as the basis for initial segmentation. These maxima grid points are `seeds' of structures, and we expand them in an iterative process to obtain the final structures. During each iteration, we assess all connected grid points to ascertain whether any are lower than their connected structure grid points but higher than zero. We then expand these structures to include neighboring grid points that meet the aforementioned criteria. The expansion of structures continues until no neighboring grid points can be incorporated. At that point, our segmentation process is complete.

As mentioned in Section~\ref{sec:cs-locmx}, a single structure may manifest as multiple islands across different scales. Therefore, it is necessary to associate each halo candidate maximum with an island to avoid duplicate detections. Furthermore, it is crucial to exclude substructures, as they might also manifest as local maxima and islands, which can lead to confusion when identifying halos. We zoom in on all the islands and their corresponding background sea at different scale parameters to the highest resolution, which is the resolution of the largest scale parameter $w$. Every halo candidate is then assessed from smaller to larger scale parameters. If a halo candidate is located within an island at its scale and this island does not overlap with any confirmed structures, it is registered as a halo.

\begin{figure*}[t]
    \centering
    \includegraphics[width=0.95\textwidth]{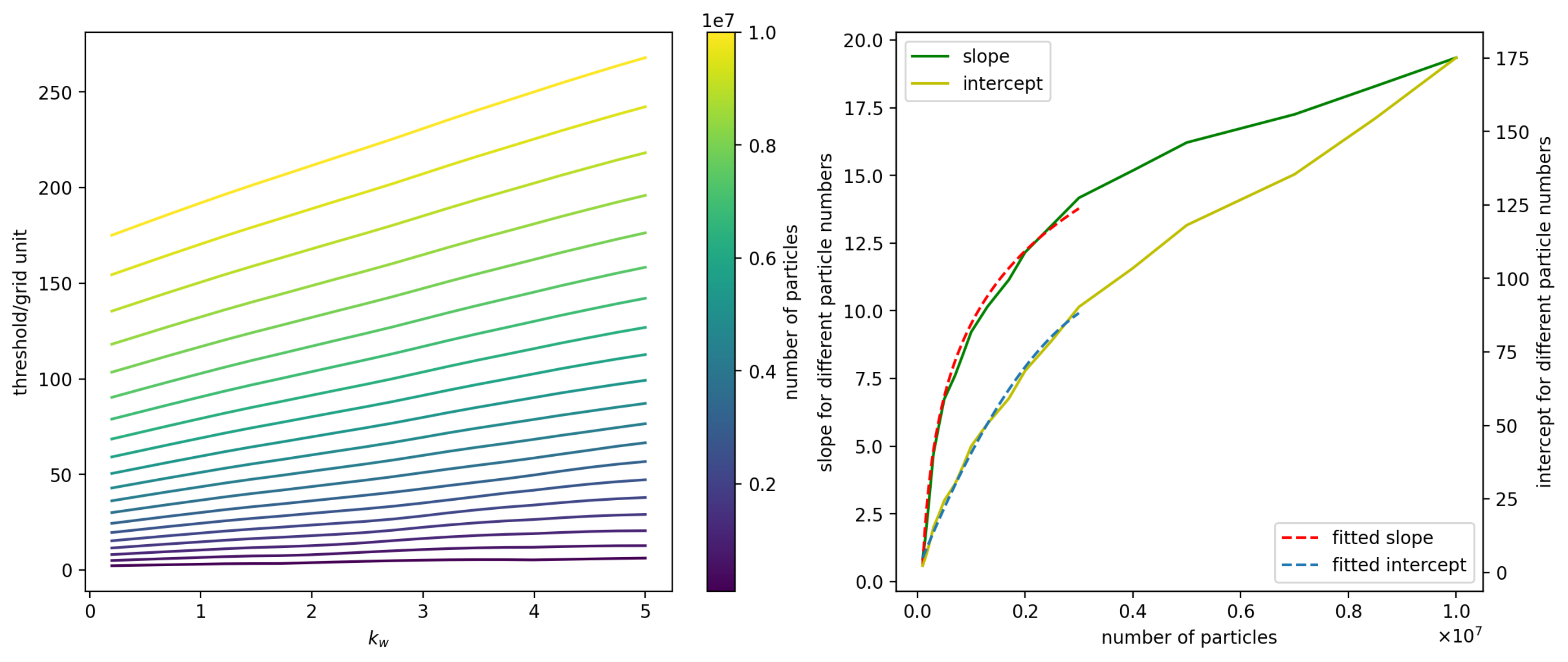}
    \caption{Left: threshold-$k_w$ relationship at different particle numbers, ranging from $N = 10^5$ (bottom, black) to $10^7$ (top, yellow). We can observe that the relationship between the variables demonstrates high linearity. Right: the slope and intercept of the linear relation from the left panel are shown as functions of the particle number. The dotted lines represent the empirical formulas fitted to the data.}    
    \label{slope_and_intercept}
\end{figure*}

\subsection{Refinement and Smoothing of Halo Boundaries}

After registering all halos on a map with the highest resolution (smallest scale), we assess each particle to determine if it is located within a halo and label it with the corresponding halo index. However, constrained by our grid resolution, the CWT halo in the densest regions has very sharp edges. To address this issue, we propose checking if the particle is located within any particular halo grid, as shown in the left panel of Figure~\ref{Smooth}, where the color encodes the probability density of particles derived by a Gaussian kernel. To alleviate the serrated edges, we examine the halo attribution of the four nearest grid points for each particle and calculate the particle's weight to the four grid points using a CIC weighting formula, as
\begin{equation}
  q_{i,j} = (1 - {\rm x}_i)(1 - {\rm y}_j),
\end{equation}
where ${\rm x}_i$, ${\rm y}_j$ represent the normalized distances to the $i,j$ th grid point in x and y direction, respectively, and $i,j \in [1,2]$ denote the four nearest grid points. After obtaining the weights, we sum them up if some grid points are assigned to the same halo. The particle is then assigned to the halo with the largest weight (we consider the background as halo `0'). As illustrated in Figure~\ref{Smooth}, the once severe serrated edges have become noticeably smoother following the weighting process. Despite our smoothing strategy, the CWT halos still present numerous sharp edges, a consequence of the limited resolution. In the 3D case, these sharp edges can be further suppressed through an unbinding procedure as is the case with most halo finders, because we can access the full dynamical information of those particles instead of limited pseudo-2D data. Finally, we discard halos that contain too few particles and obtain our final identification result.

\section{Significance and Resolution}
\label{sec:resolution}

\subsection{Determining Significance Levels and Threshold Criteria}

In the previous section, we applied thresholding to the local maxima due to the presence of Poisson noise. While the CWT of a uniform density field should ideally be zero, the representation of the true density field using finite particles constitutes a Poisson process. This process induces fluctuations in the CWT values, resulting in the generation of spurious maxima and minima. Principally, there are two distinct methods to calculate the threshold when identifying structures using wavelets: one is based on the standard deviation of the data \citep[e.g.,][and others]{Grebenev1995, Lazzati1999, Cayon2000, Barnard2004}, and the other employs Monte Carlo simulations \citep[e.g.,][]{Bijaoui1992, Flin2006, Pagliaro1999, Kazakevich2004}. In this paper, we adopt the Monte Carlo method, which provides an empirical formula that significantly simplifies the identification process.

The concept of employing a random simulation to assess significance was initially introduced by \citet{Bijaoui1992}. This approach was subsequently adopted by \citet{Slezak1993} and \citet{Escalera1994} for the identification of structures using wavelets. The likelihood of identifying a spurious structure post-thresholding is termed the significance level. To evaluate the threshold, we can create a uniform Poisson noise ensemble that mirrors our data in particle count. It is crucial to recognize that each local maximum within the noise corresponds to a potential false structure. Consequently, following thresholding, any surviving grid points could result in a false identification. Therefore, the probability of a false identification, also referred to as the significance level, can be estimated by determining the proportion of CWT values that surpass the threshold.

\begin{figure*}[t]
    \centering
    \includegraphics[width=0.98\textwidth]{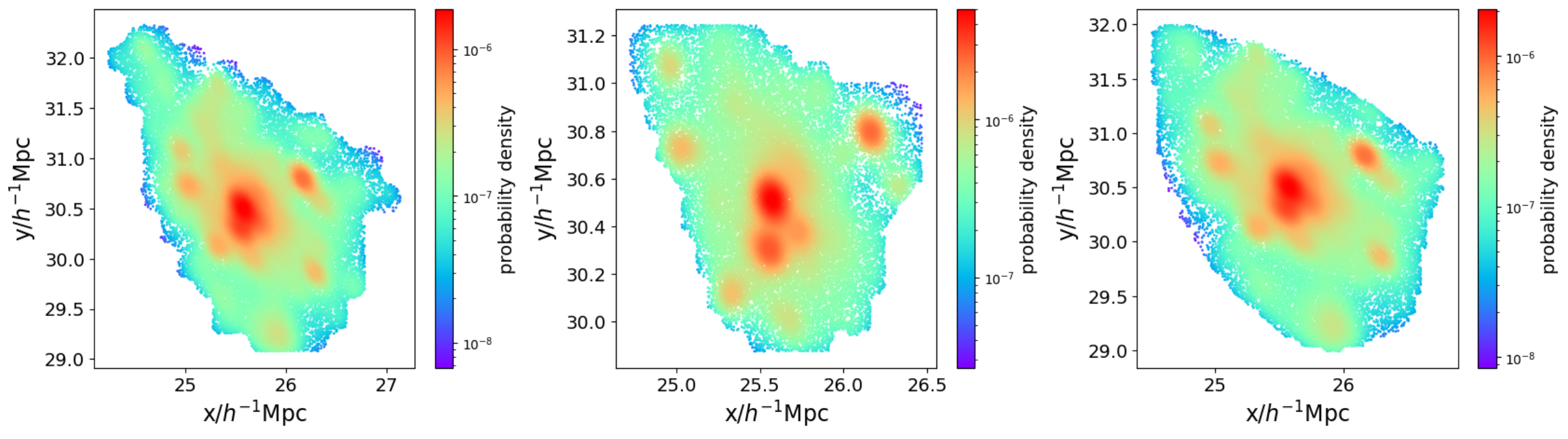}
    \caption{The largest halo within the slice at ${\rm z} = 38.0\mpch$, presented from left to right at low ($500\times k_w+250$), medium ($1000\times k_w+500$), and high ($2000\times k_w+1000$) resolutions.  The color represents the probability density of particles.} 
    \label{resolution}
\end{figure*} 

To determine these thresholds, we conducted 800 iterations of Monte Carlo simulations at each scale, averaging the thresholds at a significance level of 0.005. Running Monte Carlo simulations each time we identify halos is time-consuming and computationally intensive. The threshold is dependent solely on the dimensionless scale parameter $k_w$ and the number of particles, assuming the resolution remains unchanged. Hence, by performing Monte Carlo simulations for a specified resolution and across a wide range of values for both the particle number and $k_w$, we can derive an empirical formula. This formula will facilitate future halo identification processes.

We find that the threshold grows linearly with $k_w$ for a fixed particle number $N_{\rm p}$, as illustrated in the left panel of Figure~\ref{slope_and_intercept}. Thus, we assume our empirical formula has the following form:
\begin{equation}
    TH(k_w,N_{\rm p}) = k(N_{\rm p})\times k_w + b(N_{\rm p}),
\end{equation}
where $k$, $b$ represent the slope and intercept of the linear function, respectively, and both are solely dependent on $N_{\rm p}$. From the right panel of Figure~\ref{slope_and_intercept}, the slope and intercept can be fitted using a logarithmic function and a quadratic function, respectively. Consequently, our empirical formula takes the following form:
\begin{equation}
    TH(k_w,N_{\rm p}) = k(N_{\rm p})\times (k_w-0.2) + b(N_{\rm p}),
    \label{empirical_formula}
\end{equation}
where
\begin{flalign}
    k(N_{\rm p}) = & 8.9326 \times {\rm log}_{10}N_{\rm p} - 44.0766, \nonumber
\end{flalign}
\begin{flalign}
    b(N_{\rm p}) = & -5.4418 \times 10^{-12} N_{\rm p}^2 + 4.5586 \times 10^{-5}N_{\rm p}\nonumber\\    
    - & 4.3582.\nonumber
\end{flalign}
For improved accuracy at the low-particle end, curve fitting excludes results with particle numbers above $3\times 10^6$. Equation~(\ref{empirical_formula}) is valid within the range $0.1<k_w<5$ and $10^5<N_{\rm p}<3\times 10^6$, exhibiting an approximate dispersion of 10\%. As $k_w$ increases, the support of each particle seldom intersects, causing the threshold to decrease with larger $k_w$ and to deviate from the linear relationship in scenarios with fewer particles. 

\subsection{Optimal Grid Resolution for Accurate Halo Representation}

To accurately represent the original CWT with a finite grid, it is preferable to sample the CWT on a grid that is not too sparse, given its approximate compact support. Each grid point can only sample particles within a finite area centered on the grid point. However, increasing the density of the grid points significantly increases computational costs. Consequently, we evaluated three distinct resolution sets: 
\begin{flalign}
    &N_{\rm g,h} = {\rm Int}(2000\times k_w)+1000, \nonumber\\    
    &N_{\rm W,h} = {\rm Int}(2\times (10/k_w+20))+1, \nonumber
\end{flalign}
for high resolution; 
\begin{flalign}
    & N_{\rm g,i} = {\rm Int}(1000\times k_w) + 500, \nonumber \\    
    & N_{\rm W,i} = {\rm Int}(2\times (10/k_w+10)) + 1, \nonumber
\end{flalign}
for intermediate resolution; and finally, for low resolution, we set 
\begin{flalign}
    &N_{\rm g,l} = {\rm Int}(500\times k_w)+250, \nonumber\\    
    &N_{\rm W,l} = {\rm Int}(2\times (5/k_w+6))+1. \nonumber
\end{flalign}
We depict the same largest halo in a $38\mpch$ slice across different resolutions to illustrate the resolution effect in Figure~\ref{resolution}, with color indicating the probability density of particles. The identification yielded 1059, 1673, and 1257 halos containing 164,885, 180,834, and 183,772 particles, with computation times of 287.1, 357.4, and 1246.2 s for low, medium, and high resolutions, respectively. For the largest halo, the low and high-resolution images contain more particles (77,695 and 80,615) than the medium resolution (49,599), resulting in a lower count of halos detected in their runs. The low-resolution and high-resolution images exhibit more elongated shapes, whereas the medium-resolution image appears more compact. All three figures represent the same halo, and without a doubt, the high-resolution image provides the smoothest boundary.
\begin{figure*}[t]
    \centering
    \includegraphics[width=0.95\textwidth]{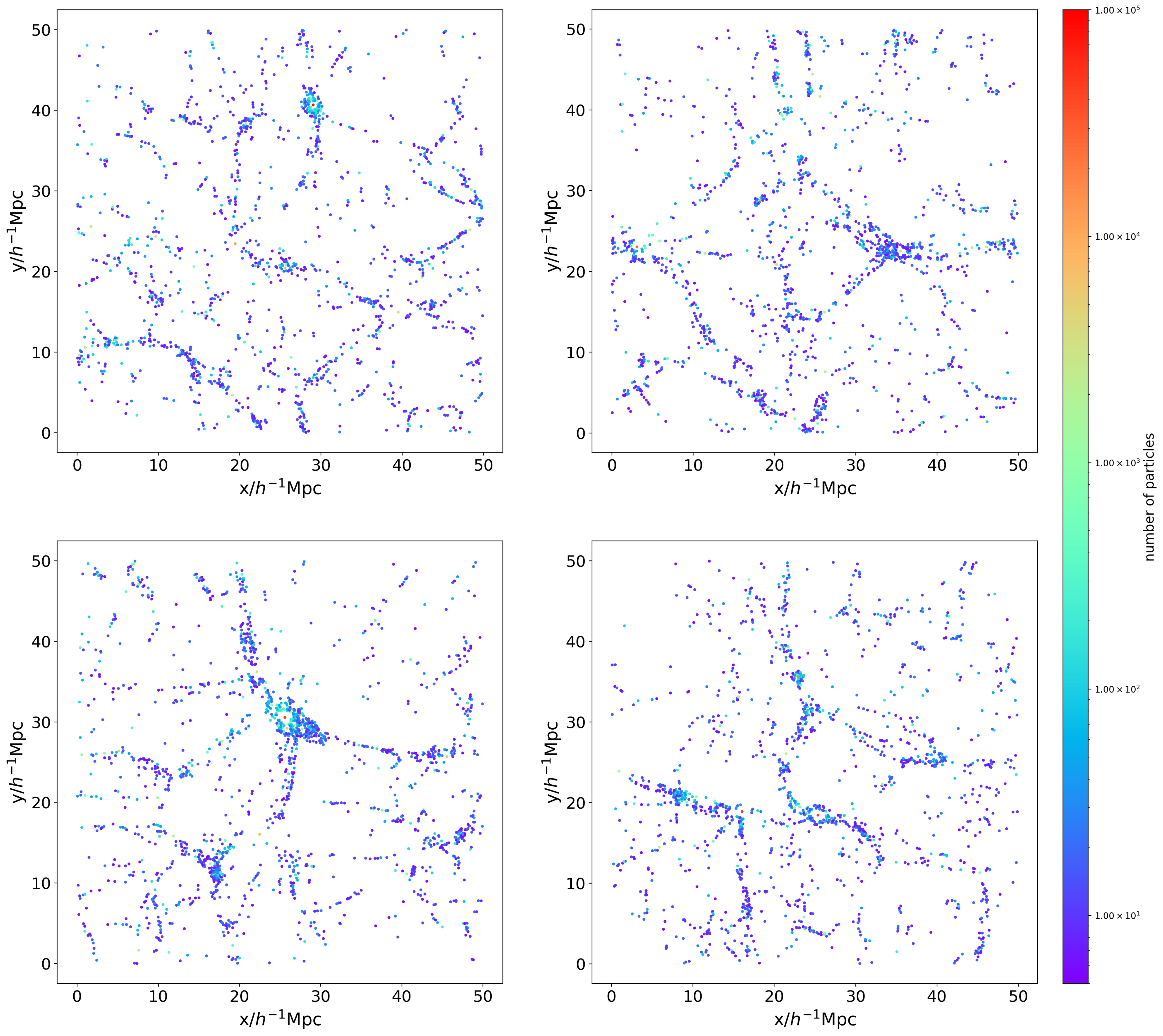}
    \caption{The CWT halos of our pseudo-2D data. The slices are arranged in the order of ${\rm z} = 11.4$, $28.4$, $38.0$, and $45.7\mpch$ from top left to bottom right.}
    \label{CWT_halos}
\end{figure*} 

As defined in Section~\ref{sec:cs-locmx}, the true structure corresponds to the largest CWT maxima across scales. So, in both high-resolution and low-resolution runs, the largest halo is incorrectly selected due to the shift of the local maxima. Our cross-scale comparison is based on the positions of these maxima. If a maximum falls outside of our comparison range, it will still be considered valid, even if it does not accurately represent the true scale of the structure. The CWT maxima value of structures shown in the left and right panels of Figure~\ref{resolution} only reaches approximately $200\mpch$, while the value of the true cross-scale CWT maxima is larger than $400\mpch$.

\begin{figure*}[t]
    \centering
    \includegraphics[width=0.95\textwidth]{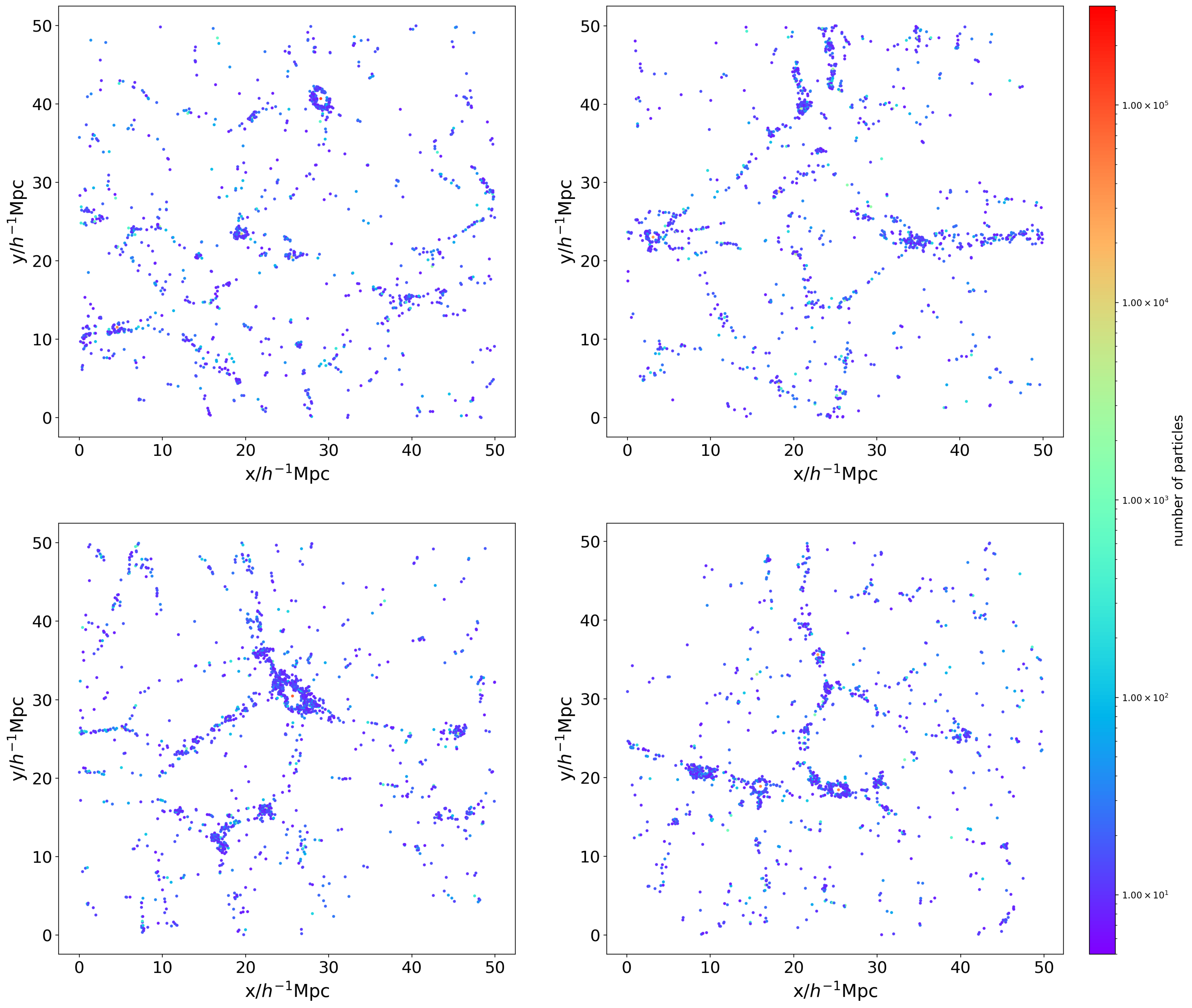}
    \caption{Same as Figure~\ref{CWT_halos}, but these halos are derived using the FOF method.}    
    \label{FOF_halos}
\end{figure*}

\begin{figure*}[t]
    \centering
    \includegraphics[width=0.98\textwidth]{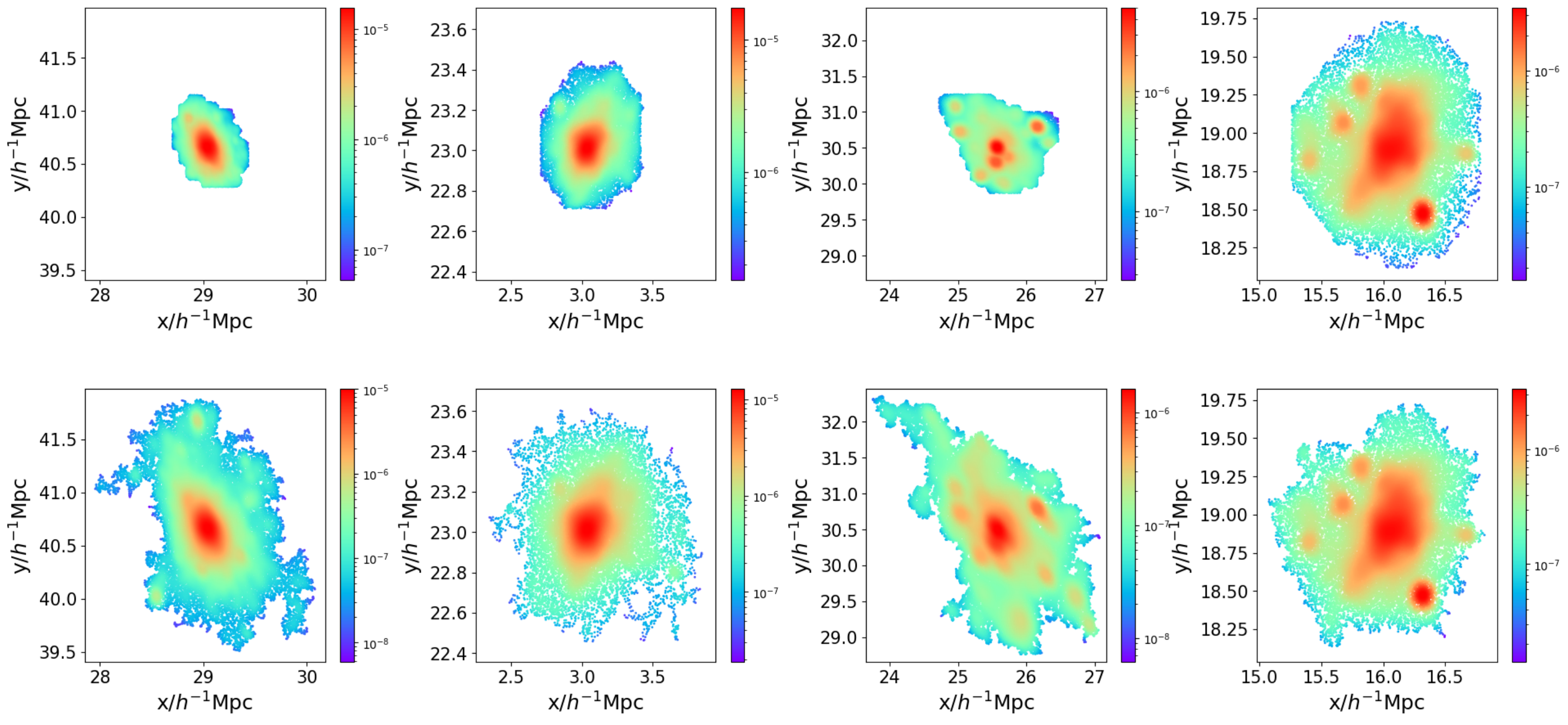}
    \caption{The largest halos in Table~\ref{tab:property} identified by the CWT (top) and the FOF (bottom) methods, arranged in the order of ${\rm z} = 11.4$, $28.4$, $38.0$, and $45.7\mpch$ from left to right. All particles within these halos are scattered and colored based on the probability density of particles.}    
    \label{largest_halo}
\end{figure*}

\begin{figure*}[t]
    \centering
    \includegraphics[width=0.98\textwidth]{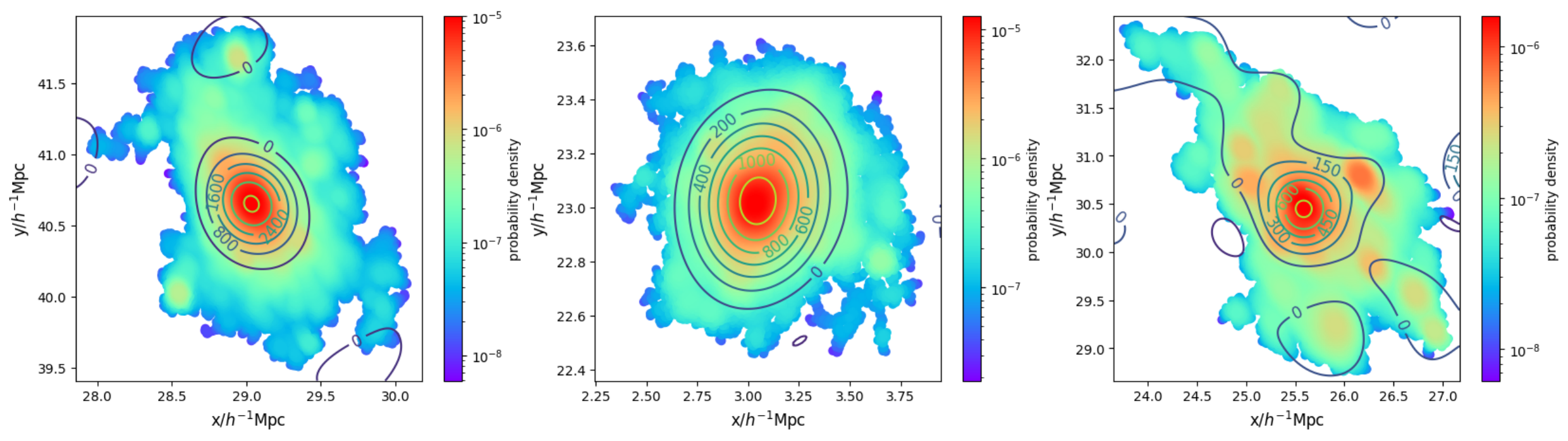}
    \caption{The three largest FOF halos from the slices ${\rm z} = 11.4, 28.4$, and $38\mpch$ are shown, along with their CWT contours. The color represents the probability density of particles. In dense regions, the CWT halo excludes the outermost parts of the FOF halo.}   
    \label{contour}
\end{figure*}
Increasing the grid resolution typically separates closely linked structures that are obscured by low spatial sampling. However, in this particular case, doubling the resolution leads to the inclusion of even more particles in the largest halo, which is comparable to a FOF cluster containing 86,331 particles. Moreover, the shift of maxima results in the selection of larger structures as halos, leading to a reduction in the number of halos and the loss of certain halos. For example, the fifth-largest halo in the $z=11.4 \mpch$ slice, the third-largest halo in the $z=28.4 \mpch$ slice, and the second- and sixth-largest halos in the $z=38.0 \mpch$ slice are no longer identified. Given that the introduction of the CWT boundary condition is expected to mitigate the `linking bridges' problem inherent in the FOF method, it is anticipated that the CWT catalog will include a greater number of low-to-medium mass halos (15-3000 particles). In this range, the FOF method identifies 400–500 halos, while our intermediate-resolution result identifies approximately 500 halos. In contrast to our objective, both high- and low-resolution results exhibit a reduction in the number of halos in comparison to that of the FOF method. 

To enrich the halo catalogs and minimize computational costs, we ultimately selected an intermediate resolution. At this resolution, we sample each individual CWT - specifically, just the MH wavelet itself - using more than 400 grid points within its support. This approach allows for an accurate representation of the CWT's shape. 
\begin{deluxetable}{lccc}
\renewcommand{\arraystretch}{1.2}
\tablehead{
\colhead{Halo Source} & \colhead{Total Particles} & \colhead{Halo Number} & \colhead{Largest Halo Size}
}
\tablecaption{The properties of FOF halos and CWT halos in our four slices.}
\tablenum{2}
\startdata
        FOF114    & 233,211         & 1340       & 119,445            \\
        CWT114    & 209,959         & 1494       & 89,814             \\
    \hline
        FOF284    & 142,777         & 1318       & 33,133             \\ 
		CWT284    & 128,030         & 1492       & 27,692             \\ 
    \hline
        FOF380    & 196,385         & 1801       & 86,331             \\
  		CWT380    & 180,834         & 1673       & 49,599             \\ 
    \hline
        FOF457    & 221,206         & 1489       & 47,061             \\ 
		CWT457    & 202,535         & 1545       & 46,596             \\ 
\enddata
\tablecomments{In `Halo source', FOF114 and CWT114 denote the FOF and CWT halos, respectively, from the slice at ${\rm z} = 11.4\mpch$. Other data are named in a similar fashion. `Total particles' indicates the overall number of particles within halos. `Halo number' signifies the count of halos listed in these catalogs. `Largest halo size' refers to the number of particles contained within the largest halo.}
\label{tab:property}
\end{deluxetable}
\begin{figure*}[t]
    \centering
    \includegraphics[width=0.98\textwidth]{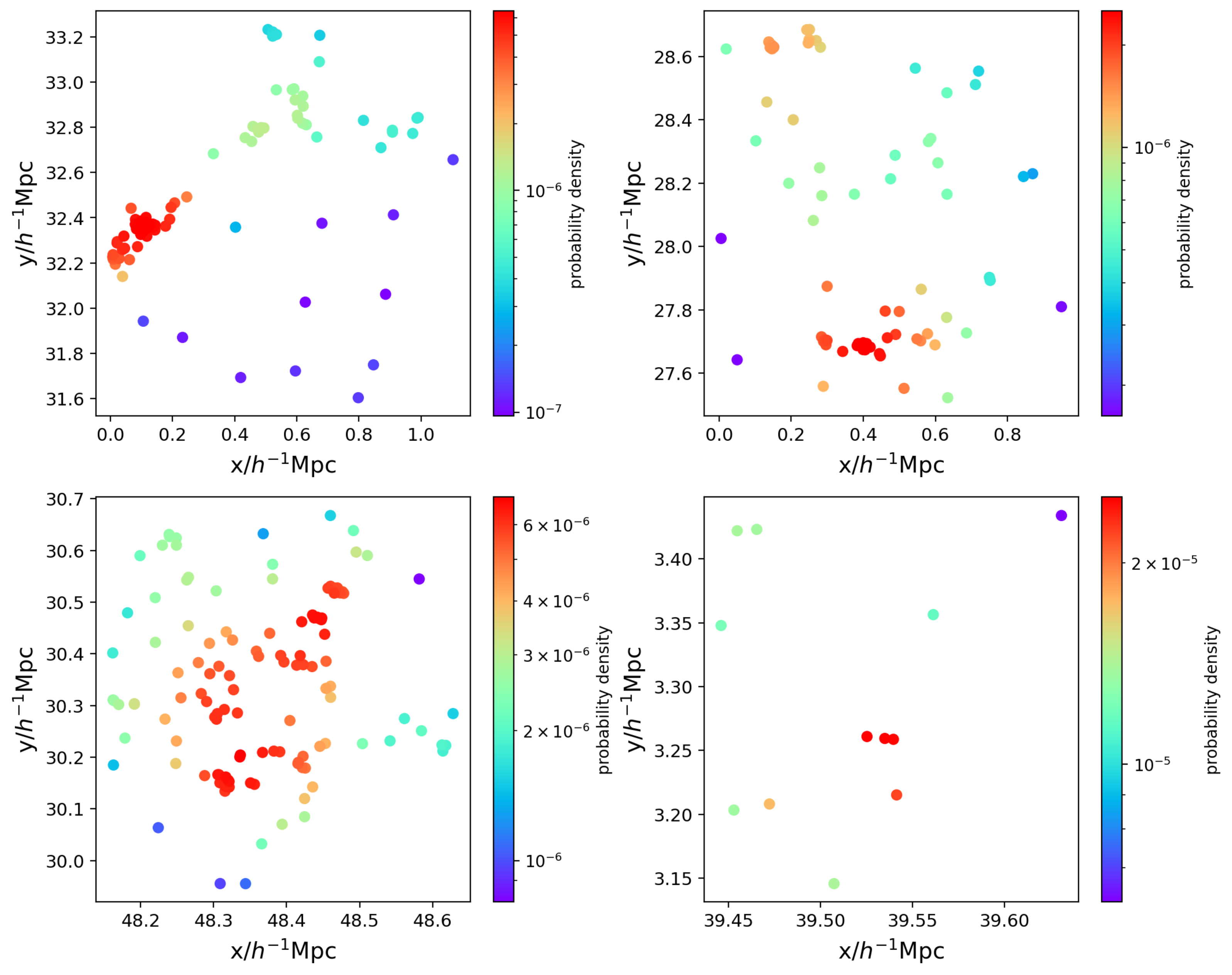}
    \caption{Four individual CWT halos from the slice of ${\rm z} = 38.0\mpch$, and the color represents the probability density of particles. The linking length in this slice is $0.018\mpch$. Consequently, these halos must contain more particles than their FOF counterparts, given that the distance between some groups of particles exceeds the linking length.}
    \label{individual_halo}
\end{figure*}

\section{Results}
\label{sec:result}
Figures~\ref{CWT_halos} and \ref{FOF_halos} present the distribution of our CWT halos and their corresponding FOF halos. The dimensionless scale parameter $k_w$ ranges from 0.05 to 2, corresponding to a physical scale of 1$\mpch$ to 0.025$\mpch$ when identifying halos. The structures of the FOF and CWT halos are similar and consistent with the CIC grid. However, the CWT halo exhibits a significantly more abundant structure. Additionally, there are more CWT halos in the medium mass range, containing $10 - 100$ particles (represented by cyan to baby blue in the color map), than there are FOF halos. For the smallest halos (represented by blue to purple in the color map), FOF halos are predominantly found in a higher mass range of approximately 10 particles (dark blue in the color map), whereas the CWT halos encompass a smaller number of particles (purple in the color map). Furthermore, isolated red points in dense regions indicate the presence of very large halos in both FOF and CWT methods.

The detailed comparisons between the FOF halos and the CWT halos are summarized in Table~\ref{tab:property}. The FOF method accounts for approximately 70\% or less of the total particles, whereas the CWT accounts for over 60\% of the particles. Specifically, the CWT method identifies halos containing 10\% fewer particles in total and yields a slightly higher number of halos compared to the FOF method. Moreover, the largest FOF halo contains approximately 30\% more particles on average than the largest CWT halo.

When focusing on the densest regions, it is evident that the largest CWT halos are consistently surrounded by several intermediate-mass halos containing a few hundred particles. In contrast, the neighborhood of the largest FOF halo contains fewer particles, even in the final slice, where the largest CWT and FOF halos are nearly identical. This discrepancy is primarily due to the compact shape of the CWT halo, although it is likely that the `linking bridge' problem inherent in the FOF method also contributes. In dense regions, the high particle density promotes the formation of a chain of particles that links two nearby halos, resulting in a larger halo.

Figure~\ref{largest_halo} presents the largest halos identified by the FOF method and their corresponding CWT halos. Each point represents a particle within this halo, and the color represents the probability density of the particles. These halos share similar positions, shapes, and even particle content, suggesting that they can be considered as the same halo classified by different methods. It should be noted that in the first three slices, the CWT halos are also identified as the largest halos. In the final slice, the largest FOF halo corresponds to the second-largest CWT halo because it contains a similar number of particles.

\begin{figure*}[t]
    \centering
    \includegraphics[width=0.98\textwidth]{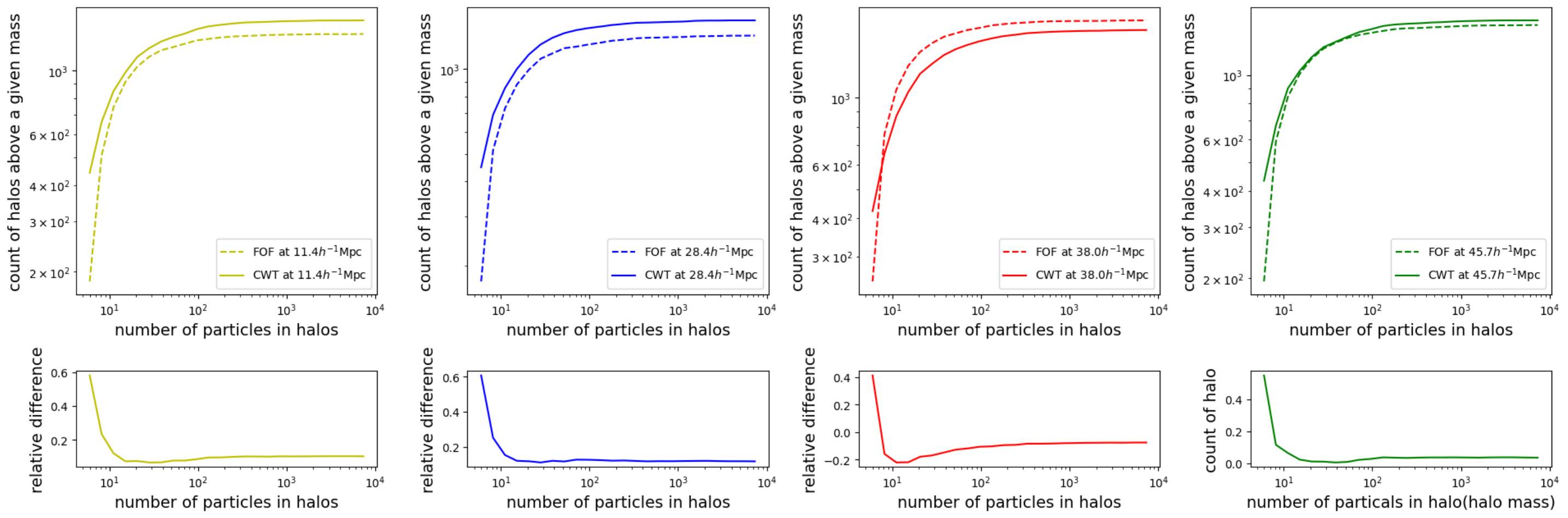}
    \caption{Top: the cumulative halo mass function across the four slices mentioned in Table~\ref{tab:property}, the solid line represents the cumulative mass function of the CWT halos, and the dashed line represents that of the FOF halos. Bottom: the relative difference between the cumulative halo mass function of FOF and CWT.}
    \label{halo_mass_func}
\end{figure*}

In the comparison between CWT and FOF halos, it is observed that CWT halos have fewer particles, are more compact, and are defined by a clearer boundary at grid points where the CWT values are less than zero and decreasing from their local maximum. On the other hand, FOF halos are more diffuse, bounded by voids devoid of particles, and contain more particles. Both FOF and CWT halos feature several density peaks (represented by red particles), revealing a multimodal structure indicative of the subhalos within them.

The local density of CWT halos spans approximately two orders of magnitude, slightly less than that of FOF halos, whose local density spans $\lesssim 3$ orders of magnitude. This discrepancy is due to the no-particle area at the boundaries of FOF halos, which extends at least the linking length. In contrast, the boundary of CWT halos is defined by the change in the density field. As previously mentioned, the CWT of a constant background is zero. Therefore, our CWT halos are delineated solely based on the variation of the density field, resulting in sharp boundaries. This enables us to distinguish between different structures even in highly dense regions, identify each local density peak as a distinct halo, and effectively mitigate the linking bridge problem.

\begin{figure}[t]
    \centering
    \includegraphics[width=0.44\textwidth]{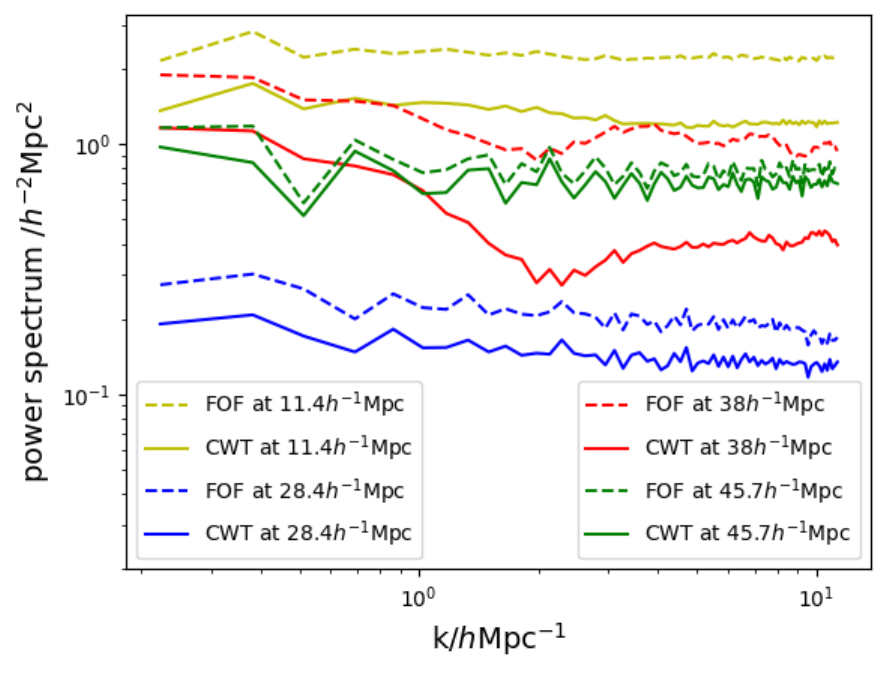}
    \caption{Power spectrum of four slices: the solid line represents the power spectrum of the CWT halos, while the dashed line represents that of the FOF halos. It is evident that the CWT halos have a lower power spectrum than their FOF counterparts at all wavenumbers.}  
    \label{ps}
\end{figure}

All four CWT halos exclude the outermost layer of their corresponding FOF halos, demonstrating a marginally more compact density distribution. The compact shape of the CWT halo is attributed to the boundary condition we use. CWT equals zero implies that the convolution of the negative part of the wavelet with the central density peak is comparable to the convolution of the positive part of the wavelet with the outer part of the density peak. The high-density peak forces the boundary where CWT equals zero inward to alleviate the large negative convolution. Consequently, the diffuse and extended outermost parts are excluded from the CWT halos. Figure~\ref{contour} displays the largest FOF halos and their corresponding CWT contours. The scale of the CWT for these contours is set to $w \sim 4\hmpc$ for the first two halos and $w \sim 3\hmpc$ for the last halo, consistent with the scales of their respective CWT halos.

The contours represent the boundary of the CWT halo with a high enough resolution that cannot be achieved when running the global halo identification program. The CWT halo exhibits a smooth boundary, contrasting with the fractal structure observed in FOF when the resolution is sufficiently high. Additionally, we affirm the importance of introducing our new boundary condition that delineates adjacent structures at the saddle point between them, as depicted in the final image of Figure~\ref{contour}. The extensive support of the wavelet at a large scale could create a positive bridge linking two adjacent halos if we were to use `CWT equals zero' as the sole boundary condition. Moreover, it can be seen that with a sufficiently high grid resolution, the boundary of the intermediate-resolution image most closely matches the true boundary.

In low-density regions, the finite resolution no longer poses a problem due to the sharp boundary, as the grid points that constitute the halo already encompass the no-particle voids. This implies that the uniform background density is close to zero in low-density regions. Figure~\ref{individual_halo} illustrates examples of these CWT halos. Within these halos, the distance between some groups of particles exceeds the linking length, leading to their corresponding FOF halo having fewer particles or even not forming at all due to an insufficient number of particles. This could partially account for why there are more CWT halos than FOF halos in the medium mass range.

Furthermore, the CWT method separates structures based on their CWT values, enabling the identification of relatively weaker and more extended structures within loose regions. These structures, while having a low CWT value, still exceed the threshold and are characterized by a negative ring surrounding them. Consequently, our CWT method can identify these particles as distinct individual halos.

The density distribution of halos can be further examined through the cumulative halo mass functions shown in Figure~\ref{halo_mass_func}. These functions show that the number of CWT halos exceeds that of FOF halos in the very low mass range (less than 10 particles, near the minimum particle threshold), where CWT shows a relative excess of over  50\%. However, as the mass increases to approximately 10 particles, the number of CWT halos drops sharply, creating a valley in the relative difference. In the higher mass range, the number of CWT halos gradually approaches that of FOF halos and surpasses FOF when the mass reaches approximately 30 particles, at which point the relative difference begins to grow. Beyond this threshold, this excess stabilizes at approximately 20\%, with a dispersion of about 20\%.

Overall, the cumulative halo mass functions exhibit a decelerating growth rate as mass increases, indicating a declining trend in the mass function. The abrupt decline observed in the relative difference around 10 particles demonstrates a high degree of consistency with the scatter observed among those. Specifically, the CWT method indicates a higher abundance in the intermediate and lower mass ranges.

Finally, we compute the halo power spectrum for four slices by assigning the halos to a $256^2$ grid using a weighted CIC method. The resulting power spectra are displayed in Figure~\ref{ps}. It is observed that the FOF halos exhibit higher amplitude across all wavenumbers due to containing more particles than the CWT halos. There is a high degree of consistency in the power spectrum across all scales. After a simple translation, the power spectra of the CWT and FOF halos nearly coincide. In summary, the clustering intensity of the CWT halos is similar to, but slightly less than, that of the FOF halos at every scale.

\section{Conclusions and Discussions}
\label{sec:concl}

In this paper, we employ a wavelet-based algorithm to identify halos within the pseudo-2D data generated from the SIMBA simulation and compare the results with the traditional FOF method. Our algorithm leverages the ability to discern structures across both space and scale via the CWT, which can be considered as a `mathematical microscope' for delving into the intricate structure of the density field and automatically extracting its clustering properties. To locate the halos in position-scale space, we initially select a range of scale parameters $w$ and compute the grid CWT for a series of $w_i$ within this range. We then identify the local maxima of the CWT at each scale and compare them across scales to retain the most prominent maximum at each position. Once all local maxima have been detected, we segment the CWT grid into distinct islands to delineate the shape of these local maxima. The islands are zoomed and projected back into the simulation space in descending order of scale, discarding any overlapping islands originating from smaller scales. Subsequently, each particle is assessed to ascertain whether it is part of an island. Halo particles are designated, and only halos comprising more than five particles are retained.

To prevent the incorrect detection of Poisson noise as halos, we conduct Monte Carlo simulations to assess the significance of each local maximum. By introducing a random set of mock particles, equal in number to the actual data, into the simulation space, we can establish a detection threshold and its significance level. This is done by analyzing the histogram of the CWT values obtained from the mock data. To eliminate the time-consuming aspect of Monte Carlo calculations, we prerun simulations across a broad range of particle numbers and scales. The outcomes are then fitted to an empirical formula. This formula allows us to set a threshold at the 99.5\% significance level with an error margin of approximately 10\%. 

Furthermore, we examine the effect of resolution on halo detection. Low resolution can make it difficult to distinguish between closely located halos, potentially reducing the number of detected halos to two-thirds of what would be found at standard resolution. On the other hand, doubling the resolution provides a more detailed definition of halo boundaries. However, this increased resolution might not accurately represent the true halo structure due to shifts in the location of maxima, and it demands a computational cost four times greater than that of the standard resolution.

Employing our method, we have successfully compiled a halo catalog. The properties of our CWT halos are compared with those of the FOF halos and are summarized below:
\begin{enumerate}
  \item The spatial distribution of the CWT halos is similar to that of the FOF halos, and both exhibit a strong correlation with the density field. They demonstrate high number densities in dense regions, show no detection in voids, and even form a web-like structure across all four slices.

  \item The CWT method identifies a greater number of halos compared to the FOF method, even though it contains fewer total particles. Specifically, the FOF method accounts for approximately 70\% of the total particles, whereas the CWT method includes approximately 10\% fewer particles. Furthermore, the CWT method yields a slightly higher count of halos than the FOF method.

  \item The boundary of a CWT halo is delineated by the grid points where the CWT value is less than zero and exhibits a decreasing trend. This boundary serves as a natural demarcation when attempting to partition the entire space into distinct structures. However, owing to our finite grid resolution, the boundary appears comparatively sharp as opposed to the identifications obtained from high-resolution data.

  \item Based on the cumulative halo mass function, it can be inferred that in the lowest mass range (around five particles), the CWT halo shows the greatest excess in the number of halos and then drops sharply around 10 particles. As the halo mass grows larger, the number of CWT halos gradually increases to match that of the FOF, and ultimately stabilizes at an excess of approximately 20\%.

  \item Despite having a lower amplitude, the halo power spectrum of the CWT halos is consistent with that of the FOF halos across all scales. This indicates that while the clustering intensity of CWT halos may be less pronounced than that of FOF halos, the overall trends they follow are similar.
\end{enumerate}

Overall, our findings indicate that the CWT is capable of identifying halos in cosmological simulations, owing to its high consistency with the FOF method. The aim of this paper is to demonstrate the feasibility of our CWT approach. It should be noted that there is significant potential for improvement in both our program and algorithm, as discussed in the Appendix~\ref{sec:performance}. Additionally, our algorithm relies solely on the positional data of these particles. To ascertain that any group of particles constitutes a halo that more closely resembles reality, we require the complete dynamic information of the particles. This is because dark matter halos are gravity-bound, virialized structures. Consequently, in future work, we will incorporate the velocity information of particles to achieve a more accurate representation of the dynamical structure of our CWT-identified halos.

In the current work, we chose to work with 2D rather than 3D data for several reasons as follows. (1) Compared with the 3D case, halos identified with 2D data can be completed with lower costs in processing time, memory, and storage space, making the development and testing of algorithms more efficient. (2) Working in the 2D case can simplify the problem, allowing us to more easily identify and address issues without the added complexity of an additional dimension. (3) 2D data is easier to visualize, which is very helpful for demonstrating and understanding the effects of an algorithm. (4) Working in the 2D case can serve as an intermediate step toward transitioning to the 3D case; the concept can first be proven in 2D before extending the research to the 3D case. Despite these benefits, our 2D test is not yet practical for real-world applications, given that we inhabit a 3D universe. To develop a program capable of analyzing real data rather than pseudo-2D data, we must account for the 3D case and optimize our program for enhanced performance. 

Currently, the performance of our program incurs unacceptable time costs when processing significantly larger 3D datasets, necessitating a fast wavelet algorithm for 3D particle data \citep[e.g.,][]{Romeo2008, Wang2023}, which will be the target of our next study. Furthermore, the capability to discern structures across scales using CWT holds the potential to detect substructures within CWT halos. This capability presents an exciting avenue for exploration in our future research.

\section*{Acknowledgments}

The authors thank the anonymous referee for helpful comments and suggestions. M.X.L. thanks Prof. Pin Lyu for his assistance with some numerical techniques. The authors also express their gratitude to the SIMBA team for making their data publicly available.

\software{NumPy \citep{vanderWalt2011,Harris2020}\footnote{\url{https://numpy.org/}},  SciPy \citep{Virtanen2020}\footnote{\url{https://scipy.org/}}, Matplotlib \citep{Hunter2007}\footnote{\url{https://matplotlib.org/}}, Jupyter Notebook\footnote{\url{https://jupyter.org/}}, yt project \citep{Turk2011}\footnote{\url{https://github.com/yt-project}}}

\appendix
\restartappendixnumbering
\section{Performance Evaluation of the CWT Method}
\label{sec:performance}

\begin{figure}[t]
    \centering
    \includegraphics[width=0.44\textwidth]{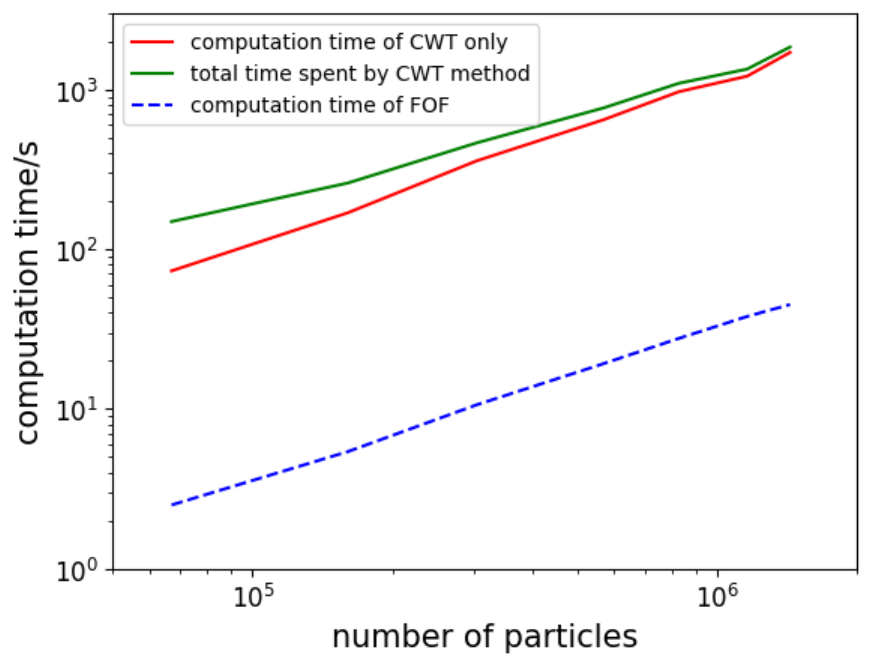}
    \caption{The computation time varies with different data sizes, ranging from 66,942 to 1,435,111 (which corresponds to a thickness from $0.02\mpch$ to $0.5\mpch$). The blue dashed line represents the computation time for the FOF method, while the solid line represents that for the CWT method, both measured in seconds. The $\mathcal{O}(N\log N)$ time complexity for the FOF method is exhibited as linear in this log-log plot.}
    \label{time}
\end{figure}

To properly evaluate the performance of an algorithm, it is crucial to understand its actual capabilities. For our CWT method, we conducted tests to measure the time required for calculating data of varying sizes. We select slices with thicknesses ranging from $0.02\mpch$ to $0.5\mpch$ to generate test pseudo-2D data for our halo identification program. The tests are conducted on a platform equipped with an AMD Ryzen 9 7900X CPU and 64GB of DDR5 6000 MHz RAM.

The program consists of two components: calculating grid CWT and characterizing local maxima. We test the time costs for each part individually and compare them with those of the FOF method. The results are depicted in Figure~\ref{time}. Both the FOF method, as calculated by the yt project, and our CWT method exhibit linear time complexity in our tests. The time cost for the former part of our program (represented by the red line) increases linearly with data size, while the latter part demonstrates significantly less variation. This is because, even with fewer particles in the thinnest slice, the number of local maxima does not change rapidly. The slice of $0.02\mpch$ contains 2,481 local maxima, and concurrently, the thickest slice of $0.5\mpch$ contains only 4270; hence, the time required to characterize them varies correspondingly slowly.

The performance of our program is significantly inferior to that of the yt project. We spend 60 times more computation time on the smallest data set and 40 times more on the largest data set. However, it is not advisable to compare the performance of a test program directly with that of a well-optimized project. Therefore, our primary goal is to confirm the $\mathcal{O}(N)$ time complexity of our program.

\bibliography{Ref}{}

\begin{thebibliography}{}
\expandafter\ifx\csname natexlab\endcsname\relax\def\natexlab#1{#1}\fi
\providecommand{\url}[1]{\href{#1}{#1}}
\providecommand{\dodoi}[1]{doi:~\href{http://doi.org/#1}{\nolinkurl{#1}}}
\providecommand{\doeprint}[1]{\href{http://ascl.net/#1}{\nolinkurl{http://ascl.net/#1}}}
\providecommand{\doarXiv}[1]{\href{https://arxiv.org/abs/#1}{\nolinkurl{https://arxiv.org/abs/#1}}}

\bibitem[{{Arnalte-Mur} {et~al.}(2012){Arnalte-Mur}, {Labatie}, {Clerc}, {Mart{\'\i}nez}, {Starck}, {Lachi{\`e}ze-Rey}, {Saar}, \& {Paredes}}]{Arnalte-Mur2012}
{Arnalte-Mur}, P., {Labatie}, A., {Clerc}, N., {et~al.} 2012, \aap, 542, A34, \dodoi{10.1051/0004-6361/201118017}

\bibitem[{{Baluev} \& {Rodionov}(2020)}]{Baluev2020}
{Baluev}, R.~V., \& {Rodionov}, E.~I. 2020, Celestial Mechanics and Dynamical Astronomy, 132, 34, \dodoi{10.1007/s10569-020-09976-2}

\bibitem[{{Barnard} {et~al.}(2004){Barnard}, {Vielva}, {Pierce-Price}, {Blain}, {Barreiro}, {Richer}, \& {Qualtrough}}]{Barnard2004}
{Barnard}, V.~E., {Vielva}, P., {Pierce-Price}, D.~P.~I., {et~al.} 2004, \mnras, 352, 961, \dodoi{10.1111/j.1365-2966.2004.07985.x}

\bibitem[{{Behroozi} {et~al.}(2013){Behroozi}, {Wechsler}, \& {Wu}}]{Behroozi2013}
{Behroozi}, P.~S., {Wechsler}, R.~H., \& {Wu}, H.-Y. 2013, \apj, 762, 109, \dodoi{10.1088/0004-637X/762/2/109}

\bibitem[{{Bendjoya} {et~al.}(1991){Bendjoya}, {Slezak}, \& {Froeschle}}]{Bendjoya1991}
{Bendjoya}, P., {Slezak}, E., \& {Froeschle}, C. 1991, \aap, 251, 312

\bibitem[{{Bijaoui} \& {Giudicelli}(1990)}]{Bijaoui1991}
{Bijaoui}, A., \& {Giudicelli}, M. 1990, Experimental Astronomy, 1, 347, \dodoi{10.1007/BF00426718}

\bibitem[{{Bijaoui} {et~al.}(1992){Bijaoui}, {Slezak}, \& {Mars}}]{Bijaoui1992}
{Bijaoui}, A., {Slezak}, E., \& {Mars}, G. 1992, in Distribution of Matter in the Universe, ed. G.~A. {Mamon} \& D.~{Gerbal}, 323--332

\bibitem[{{Cay{\'o}n} {et~al.}(2000){Cay{\'o}n}, {Sanz}, {Barreiro}, {Mart{\'\i}nez-Gonz{\'a}lez}, {Vielva}, {Toffolatti}, {Silk}, {Diego}, \& {Arg{\"u}eso}}]{Cayon2000}
{Cay{\'o}n}, L., {Sanz}, J.~L., {Barreiro}, R.~B., {et~al.} 2000, \mnras, 315, 757, \dodoi{10.1046/j.1365-8711.2000.03462.x}

\bibitem[{{Chereul} {et~al.}(1998){Chereul}, {Creze}, \& {Bienayme}}]{Chereul1998}
{Chereul}, E., {Creze}, M., \& {Bienayme}, O. 1998, \aap, 340, 384, \dodoi{10.48550/arXiv.astro-ph/9809263}

\bibitem[{{Ciprini} {et~al.}(2007){Ciprini}, {Tosti}, {Marcucci}, {Cecchi}, {Discepoli}, {Bonamente}, {Germani}, {Impiombato}, {Lubrano}, \& {Pepe}}]{Ciprini2007}
{Ciprini}, S., {Tosti}, G., {Marcucci}, F., {et~al.} 2007, in American Institute of Physics Conference Series, Vol. 921, The First GLAST Symposium, ed. S.~{Ritz}, P.~{Michelson}, \& C.~A. {Meegan}, 546--547

\bibitem[{{Daubechies}(1992)}]{Daubechies1992}
{Daubechies}, I. 1992, {Ten Lectures on Wavelets} (Philadelphia, PA: SIAM)

\bibitem[{{Dav{\'e}} {et~al.}(2019){Dav{\'e}}, {Angl{\'e}s-Alc{\'a}zar}, {Narayanan}, {Li}, {Rafieferantsoa}, \& {Appleby}}]{Dave2019}
{Dav{\'e}}, R., {Angl{\'e}s-Alc{\'a}zar}, D., {Narayanan}, D., {et~al.} 2019, \mnras, 486, 2827, \dodoi{10.1093/mnras/stz937}

\bibitem[{{Davis} {et~al.}(1985){Davis}, {Efstathiou}, {Frenk}, \& {White}}]{Davis1985}
{Davis}, M., {Efstathiou}, G., {Frenk}, C.~S., \& {White}, S.~D.~M. 1985, \apj, 292, 371, \dodoi{10.1086/163168}

\bibitem[{{Djafer} {et~al.}(2012){Djafer}, {Irbah}, \& {Meftah}}]{Djafer2012}
{Djafer}, D., {Irbah}, A., \& {Meftah}, M. 2012, \solphys, 281, 863, \dodoi{10.1007/s11207-012-0109-3}

\bibitem[{{Einasto} {et~al.}(2011){Einasto}, {H{\"u}tsi}, {Saar}, {Suhhonenko}, {Liivam{\"a}gi}, {Einasto}, {M{\"u}ller}, {Starobinsky}, {Tago}, \& {Tempel}}]{Einasto2011}
{Einasto}, J., {H{\"u}tsi}, G., {Saar}, E., {et~al.} 2011, \aap, 531, A75, \dodoi{10.1051/0004-6361/201016070}

\bibitem[{{Eisenstein} \& {Hut}(1998)}]{Eisenstein1998}
{Eisenstein}, D.~J., \& {Hut}, P. 1998, \apj, 498, 137, \dodoi{10.1086/305535}

\bibitem[{{Ellien} {et~al.}(2021){Ellien}, {Slezak}, {Martinet}, {Durret}, {Adami}, {Gavazzi}, {Raba{\c{c}}a}, {Da Rocha}, \& {Epit{\'a}cio Pereira}}]{Ellien2021}
{Ellien}, A., {Slezak}, E., {Martinet}, N., {et~al.} 2021, \aap, 649, A38, \dodoi{10.1051/0004-6361/202038419}

\bibitem[{{Escalera} {et~al.}(1994){Escalera}, {Biviano}, {Girardi}, {Giuricin}, {Mardirossian}, {Mazure}, \& {Mezzetti}}]{Escalera1994}
{Escalera}, E., {Biviano}, A., {Girardi}, M., {et~al.} 1994, \apj, 423, 539, \dodoi{10.1086/173833}

\bibitem[{{Flin} \& {Krywult}(2006)}]{Flin2006}
{Flin}, P., \& {Krywult}, J. 2006, \aap, 450, 9, \dodoi{10.1051/0004-6361:20041635}

\bibitem[{{Freeman} {et~al.}(2002){Freeman}, {Kashyap}, {Rosner}, \& {Lamb}}]{Freeman2002}
{Freeman}, P.~E., {Kashyap}, V., {Rosner}, R., \& {Lamb}, D.~Q. 2002, \apjs, 138, 185, \dodoi{10.1086/324017}

\bibitem[{{Gonz{\'a}lez-G{\'o}mez} {et~al.}(2010){Gonz{\'a}lez-G{\'o}mez}, {Blanco-Cano}, \& {Raga}}]{Gonzalez2010}
{Gonz{\'a}lez-G{\'o}mez}, D.~I., {Blanco-Cano}, X., \& {Raga}, A.~C. 2010, Advances in Space Research, 46, 22, \dodoi{10.1016/j.asr.2010.02.022}

\bibitem[{{Grebenev} {et~al.}(1995){Grebenev}, {Forman}, {Jones}, \& {Murray}}]{Grebenev1995}
{Grebenev}, S.~A., {Forman}, W., {Jones}, C., \& {Murray}, S. 1995, \apj, 445, 607, \dodoi{10.1086/175725}

\bibitem[{{Grossmann} {et~al.}(1985){Grossmann}, {Morlet}, \& {Paul}}]{Grossmann1985}
{Grossmann}, A., {Morlet}, J., \& {Paul}, T. 1985, Journal of Mathematical Physics, 26, 2473, \dodoi{10.1063/1.526761}

\bibitem[{{Harris} {et~al.}(2020){Harris}, {Millman}, {van der Walt}, {Gommers}, {Virtanen}, {Cournapeau}, {Wieser}, {Taylor}, {Berg}, {Smith}, {Kern}, {Picus}, {Hoyer}, {van Kerkwijk}, {Brett}, {Haldane}, {del R{\'\i}o}, {Wiebe}, {Peterson}, {G{\'e}rard-Marchant}, {Sheppard}, {Reddy}, {Weckesser}, {Abbasi}, {Gohlke}, \& {Oliphant}}]{Harris2020}
{Harris}, C.~R., {Millman}, K.~J., {van der Walt}, S.~J., {et~al.} 2020, \nat, 585, 357, \dodoi{10.1038/s41586-020-2649-2}

\bibitem[{{Hayn} {et~al.}(2012){Hayn}, {Panet}, {Diament}, {Holschneider}, {Mandea}, \& {Davaille}}]{Hayn2012}
{Hayn}, M., {Panet}, I., {Diament}, M., {et~al.} 2012, Geophysical Journal International, 189, 1430, \dodoi{10.1111/j.1365-246X.2012.05455.x}

\bibitem[{{Hockney} \& {Eastwood}(1981)}]{Hock-East1981}
{Hockney}, R.~W., \& {Eastwood}, J.~W. 1981, {Computer Simulation Using Particles} (New York: McGraw-Hill)

\bibitem[{{Hopkins}(2015)}]{Hopkins2015}
{Hopkins}, P.~F. 2015, \mnras, 450, 53, \dodoi{10.1093/mnras/stv195}

\bibitem[{{Hunter}(2007)}]{Hunter2007}
{Hunter}, J.~D. 2007, Computing in Science and Engineering, 9, 90, \dodoi{10.1109/MCSE.2007.55}

\bibitem[{{Kazakevich} {et~al.}(2004){Kazakevich}, {Vityazev}, \& {Orlov}}]{Kazakevich2004}
{Kazakevich}, E., {Vityazev}, V., \& {Orlov}, V. 2004, in Astronomical Society of the Pacific Conference Series, Vol. 316, Order and Chaos in Stellar and Planetary Systems, ed. G.~G. {Byrd}, K.~V. {Kholshevnikov}, A.~A. {Myllri}, I.~I. {Nikiforov}, \& V.~V. {Orlov}, 243

\bibitem[{{Knebe} {et~al.}(2011){Knebe}, {Knollmann}, {Muldrew}, {Pearce}, {Aragon-Calvo}, {Ascasibar}, {Behroozi}, {Ceverino}, {Colombi}, {Diemand}, {Dolag}, {Falck}, {Fasel}, {Gardner}, {Gottl{\"o}ber}, {Hsu}, {Iannuzzi}, {Klypin}, {Luki{\'c}}, {Maciejewski}, {McBride}, {Neyrinck}, {Planelles}, {Potter}, {Quilis}, {Rasera}, {Read}, {Ricker}, {Roy}, {Springel}, {Stadel}, {Stinson}, {Sutter}, {Turchaninov}, {Tweed}, {Yepes}, \& {Zemp}}]{Knebe2011}
{Knebe}, A., {Knollmann}, S.~R., {Muldrew}, S.~I., {et~al.} 2011, \mnras, 415, 2293, \dodoi{10.1111/j.1365-2966.2011.18858.x}

\bibitem[{{Knebe} {et~al.}(2013){Knebe}, {Pearce}, {Lux}, {Ascasibar}, {Behroozi}, {Casado}, {Moran}, {Diemand}, {Dolag}, {Dominguez-Tenreiro}, {Elahi}, {Falck}, {Gottl{\"o}ber}, {Han}, {Klypin}, {Luki{\'c}}, {Maciejewski}, {McBride}, {Merch{\'a}n}, {Muldrew}, {Neyrinck}, {Onions}, {Planelles}, {Potter}, {Quilis}, {Rasera}, {Ricker}, {Roy}, {Ruiz}, {Sgr{\'o}}, {Springel}, {Stadel}, {Sutter}, {Tweed}, \& {Zemp}}]{Knebe2013}
{Knebe}, A., {Pearce}, F.~R., {Lux}, H., {et~al.} 2013, \mnras, 435, 1618, \dodoi{10.1093/mnras/stt1403}

\bibitem[{{Knollmann} \& {Knebe}(2009)}]{Knollmann2009}
{Knollmann}, S.~R., \& {Knebe}, A. 2009, \apjs, 182, 608, \dodoi{10.1088/0067-0049/182/2/608}

\bibitem[{{Lazzati} {et~al.}(1999){Lazzati}, {Campana}, {Rosati}, {Panzera}, \& {Tagliaferri}}]{Lazzati1999}
{Lazzati}, D., {Campana}, S., {Rosati}, P., {Panzera}, M.~R., \& {Tagliaferri}, G. 1999, \apj, 524, 414, \dodoi{10.1086/307788}

\bibitem[{{Mertens} \& {Lobanov}(2015)}]{Mertens2015}
{Mertens}, F., \& {Lobanov}, A. 2015, \aap, 574, A67, \dodoi{10.1051/0004-6361/201424566}

\bibitem[{Mo {et~al.}(2010)Mo, van~den Bosch, \& White}]{Mo2010}
Mo, H., van~den Bosch, F., \& White, S. 2010, Galaxy Formation and Evolution (Cambridge University Press), \dodoi{10.1017/CBO9780511807244}

\bibitem[{{More} {et~al.}(2011){More}, {Kravtsov}, {Dalal}, \& {Gottl{\"o}ber}}]{More2011}
{More}, S., {Kravtsov}, A.~V., {Dalal}, N., \& {Gottl{\"o}ber}, S. 2011, \apjs, 195, 4, \dodoi{10.1088/0067-0049/195/1/4}

\bibitem[{{Moretti} {et~al.}(2004){Moretti}, {Guzzo}, {Campana}, {Lazzati}, {Panzera}, {Tagliaferri}, {Arena}, {Braglia}, {Dell'Antonio}, \& {Longhetti}}]{Moretti2004}
{Moretti}, A., {Guzzo}, L., {Campana}, S., {et~al.} 2004, \aap, 428, 21, \dodoi{10.1051/0004-6361:20041326}

\bibitem[{{Nelson} {et~al.}(2019){Nelson}, {Springel}, {Pillepich}, {Rodriguez-Gomez}, {Torrey}, {Genel}, {Vogelsberger}, {Pakmor}, {Marinacci}, {Weinberger}, {Kelley}, {Lovell}, {Diemer}, \& {Hernquist}}]{Nelson2019}
{Nelson}, D., {Springel}, V., {Pillepich}, A., {et~al.} 2019, Computational Astrophysics and Cosmology, 6, 2, \dodoi{10.1186/s40668-019-0028-x}

\bibitem[{{Pagliaro} {et~al.}(1999){Pagliaro}, {Antonuccio-Delogu}, {Becciani}, \& {Gambera}}]{Pagliaro1999}
{Pagliaro}, A., {Antonuccio-Delogu}, V., {Becciani}, U., \& {Gambera}, M. 1999, \mnras, 310, 835, \dodoi{10.1046/j.1365-8711.1999.02995.x}

\bibitem[{{Patrikeev} {et~al.}(2006){Patrikeev}, {Fletcher}, {Stepanov}, {Beck}, {Berkhuijsen}, {Frick}, \& {Horellou}}]{Patrikeev2006}
{Patrikeev}, I., {Fletcher}, A., {Stepanov}, R., {et~al.} 2006, \aap, 458, 441, \dodoi{10.1051/0004-6361:20065225}

\bibitem[{Peebles(1980)}]{Peebles1980}
Peebles, P. 1980, The Large-scale Structure of the Universe, Princeton Series in Physics (Princeton University Press)

\bibitem[{{Planck Collaboration} {et~al.}(2016){Planck Collaboration}, {Ade}, {Aghanim}, {Arnaud}, {Ashdown}, {Aumont}, {Baccigalupi}, {Banday}, {Barreiro}, {Bartlett}, {Bartolo}, {Battaner}, {Battye}, {Benabed}, {Beno{\^\i}t}, {Benoit-L{\'e}vy}, {Bernard}, {Bersanelli}, {Bielewicz}, {Bock}, {Bonaldi}, {Bonavera}, {Bond}, {Borrill}, {Bouchet}, {Boulanger}, {Bucher}, {Burigana}, {Butler}, {Calabrese}, {Cardoso}, {Catalano}, {Challinor}, {Chamballu}, {Chary}, {Chiang}, {Chluba}, {Christensen}, {Church}, {Clements}, {Colombi}, {Colombo}, {Combet}, {Coulais}, {Crill}, {Curto}, {Cuttaia}, {Danese}, {Davies}, {Davis}, {de Bernardis}, {de Rosa}, {de Zotti}, {Delabrouille}, {D{\'e}sert}, {Di Valentino}, {Dickinson}, {Diego}, {Dolag}, {Dole}, {Donzelli}, {Dor{\'e}}, {Douspis}, {Ducout}, {Dunkley}, {Dupac}, {Efstathiou}, {Elsner}, {En{\ss}lin}, {Eriksen}, {Farhang}, {Fergusson}, {Finelli}, {Forni}, {Frailis}, {Fraisse}, {Franceschi}, {Frejsel}, {Galeotta}, {Galli}, {Ganga}, {Gauthier}, {Gerbino}, {Ghosh}, {Giard}, {Giraud-H{\'e}raud}, {Giusarma}, {Gjerl{\o}w}, {Gonz{\'a}lez-Nuevo}, {G{\'o}rski}, {Gratton}, {Gregorio}, {Gruppuso}, {Gudmundsson}, {Hamann}, {Hansen}, {Hanson}, {Harrison}, {Helou}, {Henrot-Versill{\'e}}, {Hern{\'a}ndez-Monteagudo}, {Herranz}, {Hildebrandt}, {Hivon}, {Hobson}, {Holmes}, {Hornstrup}, {Hovest}, {Huang}, {Huffenberger}, {Hurier}, {Jaffe}, {Jaffe}, {Jones}, {Juvela}, {Keih{\"a}nen}, {Keskitalo}, {Kisner}, {Kneissl}, {Knoche}, {Knox}, {Kunz}, {Kurki-Suonio}, {Lagache}, {L{\"a}hteenm{\"a}ki}, {Lamarre}, {Lasenby}, {Lattanzi}, {Lawrence}, {Leahy}, {Leonardi}, {Lesgourgues}, {Levrier}, {Lewis}, {Liguori}, {Lilje}, {Linden-V{\o}rnle}, {L{\'o}pez-Caniego}, {Lubin}, {Mac{\'\i}as-P{\'e}rez}, {Maggio}, {Maino}, {Mandolesi}, {Mangilli}, {Marchini}, {Maris}, {Martin}, {Martinelli}, {Mart{\'\i}nez-Gonz{\'a}lez}, {Masi}, {Matarrese}, {McGehee}, {Meinhold}, {Melchiorri}, {Melin}, {Mendes}, {Mennella}, {Migliaccio}, {Millea}, {Mitra}, {Miville-Desch{\^e}nes}, {Moneti}, {Montier}, {Morgante}, {Mortlock}, {Moss}, {Munshi}, {Murphy}, {Naselsky}, {Nati}, {Natoli}, {Netterfield}, {N{\o}rgaard-Nielsen}, {Noviello}, {Novikov}, {Novikov}, {Oxborrow}, {Paci}, {Pagano}, {Pajot}, {Paladini}, {Paoletti}, {Partridge}, {Pasian}, {Patanchon}, {Pearson}, {Perdereau}, {Perotto}, {Perrotta}, {Pettorino}, {Piacentini}, {Piat}, {Pierpaoli}, {Pietrobon}, {Plaszczynski}, {Pointecouteau}, {Polenta}, {Popa}, {Pratt}, {Pr{\'e}zeau}, {Prunet}, {Puget}, {Rachen}, {Reach}, {Rebolo}, {Reinecke}, {Remazeilles}, {Renault}, {Renzi}, {Ristorcelli}, {Rocha}, {Rosset}, {Rossetti}, {Roudier}, {Rouill{\'e} d'Orfeuil}, {Rowan-Robinson}, {Rubi{\~n}o-Mart{\'\i}n}, {Rusholme}, {Said}, {Salvatelli}, {Salvati}, {Sandri}, {Santos}, {Savelainen}, {Savini}, {Scott}, {Seiffert}, {Serra}, {Shellard}, {Spencer}, {Spinelli}, {Stolyarov}, {Stompor}, {Sudiwala}, {Sunyaev}, {Sutton}, {Suur-Uski}, {Sygnet}, {Tauber}, {Terenzi}, {Toffolatti}, {Tomasi}, {Tristram}, {Trombetti}, {Tucci}, {Tuovinen}, {T{\"u}rler}, {Umana}, {Valenziano}, {Valiviita}, {Van Tent}, {Vielva}, {Villa}, {Wade}, {Wandelt}, {Wehus}, {White}, {White}, {Wilkinson}, {Yvon}, {Zacchei}, \& {Zonca}}]{Planck2016}
{Planck Collaboration}, {Ade}, P.~A.~R., {Aghanim}, N., {et~al.} 2016, \aap, 594, A13, \dodoi{10.1051/0004-6361/201525830}

\bibitem[{{Planelles} \& {Quilis}(2010)}]{Planelles2010}
{Planelles}, S., \& {Quilis}, V. 2010, \aap, 519, A94, \dodoi{10.1051/0004-6361/201014214}

\bibitem[{{Press} \& {Schechter}(1974)}]{Press1974}
{Press}, W.~H., \& {Schechter}, P. 1974, \apj, 187, 425, \dodoi{10.1086/152650}

\bibitem[{{Romeo} {et~al.}(2008){Romeo}, {Agertz}, {Moore}, \& {Stadel}}]{Romeo2008}
{Romeo}, A.~B., {Agertz}, O., {Moore}, B., \& {Stadel}, J. 2008, \apj, 686, 1, \dodoi{10.1086/591236}

\bibitem[{{Slezak} {et~al.}(1990){Slezak}, {Bijaoui}, \& {Mars}}]{Slezak1990}
{Slezak}, E., {Bijaoui}, A., \& {Mars}, G. 1990, \aap, 227, 301

\bibitem[{{Slezak} {et~al.}(1993){Slezak}, {de Lapparent}, \& {Bijaoui}}]{Slezak1993}
{Slezak}, E., {de Lapparent}, V., \& {Bijaoui}, A. 1993, \apj, 409, 517, \dodoi{10.1086/172683}

\bibitem[{{Springel} {et~al.}(2001){Springel}, {White}, {Tormen}, \& {Kauffmann}}]{Springel2001}
{Springel}, V., {White}, S. D.~M., {Tormen}, G., \& {Kauffmann}, G. 2001, \mnras, 328, 726, \dodoi{10.1046/j.1365-8711.2001.04912.x}

\bibitem[{{Turk} {et~al.}(2011){Turk}, {Smith}, {Oishi}, {Skory}, {Skillman}, {Abel}, \& {Norman}}]{Turk2011}
{Turk}, M.~J., {Smith}, B.~D., {Oishi}, J.~S., {et~al.} 2011, \apjs, 192, 9, \dodoi{10.1088/0067-0049/192/1/9}

\bibitem[{{Vall{\'e}s-P{\'e}rez} {et~al.}(2022){Vall{\'e}s-P{\'e}rez}, {Planelles}, \& {Quilis}}]{Valles-perez2022}
{Vall{\'e}s-P{\'e}rez}, D., {Planelles}, S., \& {Quilis}, V. 2022, \aap, 664, A42, \dodoi{10.1051/0004-6361/202243712}

\bibitem[{{van der Walt} {et~al.}(2011){van der Walt}, {Colbert}, \& {Varoquaux}}]{vanderWalt2011}
{van der Walt}, S., {Colbert}, S.~C., \& {Varoquaux}, G. 2011, Computing in Science and Engineering, 13, 22, \dodoi{10.1109/MCSE.2011.37}

\bibitem[{{Vavilova} \& {Babyk}(2018)}]{Vavilova2018}
{Vavilova}, I.~B., \& {Babyk}, I.~V. 2018, Odessa Astronomical Publications, 30, 239, \dodoi{10.18524/1810-4215.2018.31.146678}

\bibitem[{{Virtanen} {et~al.}(2020){Virtanen}, {Gommers}, {Oliphant}, {Haberland}, {Reddy}, {Cournapeau}, {Burovski}, {Peterson}, {Weckesser}, {Bright}, {van der Walt}, {Brett}, {Wilson}, {Millman}, {Mayorov}, {Nelson}, {Jones}, {Kern}, {Larson}, {Carey}, {Polat}, {Feng}, {Moore}, {VanderPlas}, {Laxalde}, {Perktold}, {Cimrman}, {Henriksen}, {Quintero}, {Harris}, {Archibald}, {Ribeiro}, {Pedregosa}, {van Mulbregt}, \& {SciPy 1. 0 Contributors}}]{Virtanen2020}
{Virtanen}, P., {Gommers}, R., {Oliphant}, T.~E., {et~al.} 2020, NatMe, 17, 261, \dodoi{10.1038/s41592-019-0686-2}

\bibitem[{{Vogelsberger} {et~al.}(2020){Vogelsberger}, {Marinacci}, {Torrey}, \& {Puchwein}}]{Vogelsberger2020}
{Vogelsberger}, M., {Marinacci}, F., {Torrey}, P., \& {Puchwein}, E. 2020, NatRP, 2, 42, \dodoi{10.1038/s42254-019-0127-2}

\bibitem[{{Wang} {et~al.}(2008){Wang}, {Rowan-Robinson}, {Yamamura}, {Shibai}, {Savage}, {Oliver}, {Thomson}, {Rahman}, {Clements}, {Figueredo}, {Goto}, {Hasegawa}, {Jeong}, {Matsuura}, {M{\"u}ller}, {Nakagawa}, {Pearson}, {Serjeant}, {Shirahata}, \& {White}}]{Wang2008}
{Wang}, L., {Rowan-Robinson}, M., {Yamamura}, I., {et~al.} 2008, \mnras, 387, 601, \dodoi{10.1111/j.1365-2966.2008.13292.x}

\bibitem[{{Wang} \& {He}(2021)}]{Wang2021}
{Wang}, Y., \& {He}, P. 2021, Communications in Theoretical Physics, 73, 095402, \dodoi{10.1088/1572-9494/ac10be}

\bibitem[{{Wang} \& {He}(2022)}]{Wang2022b}
---. 2022, \apj, 934, 112, \dodoi{10.3847/1538-4357/ac7a3d}

\bibitem[{Wang \& He(2023)}]{Wang2023}
Wang, Y., \& He, P. 2023, RAST, 2, 307, \dodoi{10.1093/rasti/rzad020}

\bibitem[{{Wang} {et~al.}(2022){Wang}, {Yang}, \& {He}}]{Wang2022a}
{Wang}, Y., {Yang}, H.-Y., \& {He}, P. 2022, \apj, 934, 77, \dodoi{10.3847/1538-4357/ac752c}

\end{thebibliography}
\bibliographystyle{aasjournal}
\end{CJK*}
\end{document}